\documentstyle[aps,manuscript,prd,epsf]{revtex}
\def	\bu		{B_u^-}

\def	\bd		{\overline{B}_d^0}
\def	\Bd		{\mbox{$\bd$}}
\def	\bs		{\overline{B}_s^0}
\def	\Bs		{\mbox{$\bs$}}
\def	\lamb		{\Lambda_b^0}
\def	\Lamb		{\mbox{$\lamb$}}

\def	\kst		{\overline{K}^{*0}}
\def	\Kst		{\mbox{$\kst$}}

\def	\kstc		{K^{*-}}

\def	\kstp		{K^{*0}}

\newcommand{\Ksttokpi}	{\mbox{$\kst \to K^-\pi^+$}}

\def	\jpsi		{J/\psi}
\def	\Jpsi		{\mbox{$\jpsi$}}
\def	\Jpsitoee	{\mbox{$\jpsi \to e^+e^-$}}

\newcommand{\Phitokk}	{\mbox{$\phi \to K^+K^-$}}
\newcommand{\Dtokpi}     {\mbox{$D^{0} \to K^-\pi^+$}}

\def	\lam		{\Lambda}
\def	\Lam		{\mbox{$\lam$}}
\def	\Lamtoppi	{\mbox{$\lam \to p\pi^-$}}

\newcommand{\Btogkst}	{\mbox{$\bd \to \kst\gamma$}}
\newcommand{\Butogkst}	{\mbox{$\bu \to \kstc\gamma$}}
\newcommand{\Btogphi}	{\mbox{$\bs \to \phi\gamma$}}
\def	\Lambtoglam	{\mbox{$\lamb \to \lam\gamma$}}

\def	\edz		{eD^0}
\def	\Edz		{\mbox{$\edz$}}
\def	\edx		{e^- D^0 X}

\def	\Btoedx		{\mbox{$\overline{B} \to \edx$}}

\def	\Btojpsik	{\mbox{$\bu \to \jpsi K^-$}}

\newcommand{\BE}{\begin{eqnarray}}
\newcommand{\EE}{\end{eqnarray}}

\newcommand{\ppbar}      {\mbox{$p\bar{p}$}}

\newcommand{\Br}         {\mbox{${\cal B}$}}
\newcommand{\invpb}      {\mbox{${\rm pb^{-1}}$}}

\def	\ib		{IB}
\def	\ic		{IC}
\def	\hadem		{\mbox{HAD/EM}}

\def\r#1{\ignorespaces $^{#1}$}
\begin{document}
\draft
\title{
Search for radiative $b$-hadron decays\\
in $p\bar{p}$ collisions at $\sqrt{s}$ = 1.8 TeV}
\author{
\font\eightit=cmti8
\hfilneg
\begin{sloppypar}
\noindent
D.~Acosta,\r {13} T.~Affolder,\r {24} H.~Akimoto,\r {47}
M.~G.~Albrow,\r {12} D.~Ambrose,\r {34}   
D.~Amidei,\r {26} K.~Anikeev,\r {25} J.~Antos,\r 1 
G.~Apollinari,\r {12} T.~Arisawa,\r {47}
% ...
\end{sloppypar}
\vskip .026in
\begin{center}
(CDF Collaboration)
\end{center}
\vskip .026in
\begin{center}
\r 1  {\eightit Institute of Physics, Academia Sinica, Taipei, Taiwan 11529, 
Republic of China} \\
\r 2  {\eightit Argonne National Laboratory, Argonne, Illinois 60439} \\
\r 3  {\eightit Istituto Nazionale di Fisica Nucleare, University of Bologna,
I-40127 Bologna, Italy} \\
\r 4  {\eightit Brandeis University, Waltham, Massachusetts 02254} \\
\r 5  {\eightit University of California at Davis, Davis, California  95616} \\
\r 6  {\eightit University of California at Los Angeles, Los 
Angeles, California  90024} \\  
\r 7  {\eightit Instituto de Fisica de Cantabria, CSIC-University of Cantabria, 
39005 Santander, Spain} \\
\r 8  {\eightit Carnegie Mellon University, Pittsburgh, PA  15218} \\
\r 9  {\eightit Enrico Fermi Institute, University of Chicago, Chicago, 
Illinois 60637} \\
\r {10}  {\eightit Joint Institute for Nuclear Research, RU-141980 Dubna, Russia}
\\
\r {11} {\eightit Duke University, Durham, North Carolina  27708} \\
\r {12} {\eightit Fermi National Accelerator Laboratory, Batavia, Illinois 
60510} \\
\r {13} {\eightit University of Florida, Gainesville, Florida  32611} \\
\r {14} {\eightit Laboratori Nazionali di Frascati, Istituto Nazionale di Fisica
               Nucleare, I-00044 Frascati, Italy} \\
\r {15} {\eightit University of Geneva, CH-1211 Geneva 4, Switzerland} \\
\r {16} {\eightit Glasgow University, Glasgow G12 8QQ, United Kingdom}\\
\r {17} {\eightit Harvard University, Cambridge, Massachusetts 02138} \\
\r {18} {\eightit Hiroshima University, Higashi-Hiroshima 724, Japan} \\
\r {19} {\eightit University of Illinois, Urbana, Illinois 61801} \\
\r {20} {\eightit The Johns Hopkins University, Baltimore, Maryland 21218} \\
\r {21} {\eightit Institut f\"{u}r Experimentelle Kernphysik, 
Universit\"{a}t Karlsruhe, 76128 Karlsruhe, Germany} \\
\r {22} {\eightit Center for High Energy Physics: Kyungpook National
University, Taegu 702-701; Seoul National University, Seoul 151-742; and
SungKyunKwan University, Suwon 440-746; Korea} \\
\r {23} {\eightit High Energy Accelerator Research Organization (KEK), Tsukuba, 
Ibaraki 305, Japan} \\
\r {24} {\eightit Ernest Orlando Lawrence Berkeley National Laboratory, 
Berkeley, California 94720} \\
\r {25} {\eightit Massachusetts Institute of Technology, Cambridge,
Massachusetts  02139} \\   
\r {26} {\eightit University of Michigan, Ann Arbor, Michigan 48109} \\
\r {27} {\eightit Michigan State University, East Lansing, Michigan  48824} \\
\r {28} {\eightit University of New Mexico, Albuquerque, New Mexico 87131} \\
\r {29} {\eightit Northwestern University, Evanston, Illinois  60208} \\
\r {30} {\eightit The Ohio State University, Columbus, Ohio  43210} \\
\r {31} {\eightit Osaka City University, Osaka 588, Japan} \\
\r {32} {\eightit University of Oxford, Oxford OX1 3RH, United Kingdom} \\
\r {33} {\eightit Universita di Padova, Istituto Nazionale di Fisica 
          Nucleare, Sezione di Padova, I-35131 Padova, Italy} \\
\r {34} {\eightit University of Pennsylvania, Philadelphia, 
        Pennsylvania 19104} \\   
\r {35} {\eightit Istituto Nazionale di Fisica Nucleare, University and Scuola
               Normale Superiore of Pisa, I-56100 Pisa, Italy} \\
\r {36} {\eightit University of Pittsburgh, Pittsburgh, Pennsylvania 15260} \\
\r {37} {\eightit Purdue University, West Lafayette, Indiana 47907} \\
\r {38} {\eightit University of Rochester, Rochester, New York 14627} \\
\r {39} {\eightit Rockefeller University, New York, New York 10021} \\
\r {40} {\eightit Rutgers University, Piscataway, New Jersey 08855} \\
\r {41} {\eightit Texas A\&M University, College Station, Texas 77843} \\
\r {42} {\eightit Texas Tech University, Lubbock, Texas 79409} \\
\r {43} {\eightit Institute of Particle Physics, McGill University, Montreal,
 H3A 2T8, Canada \\ and University of Toronto, Toronto M5S 1A7, Canada} \\
\r {44} {\eightit Istituto Nazionale di Fisica Nucleare, University of Trieste/
Udine, Italy} \\
\r {45} {\eightit University of Tsukuba, Tsukuba, Ibaraki 305, Japan} \\
\r {46} {\eightit Tufts University, Medford, Massachusetts 02155} \\
\r {47} {\eightit Waseda University, Tokyo 169, Japan} \\
\r {48} {\eightit University of Wisconsin, Madison, Wisconsin 53706} \\
\r {49} {\eightit Yale University, New Haven, Connecticut 06520} \\
\r {(\ast)} {\eightit Now at University of California, Santa Barbara, 
California  93106}
\end{center}
}
\address{}
\maketitle
\begin{abstract}
We have performed a search for radiative $b$-hadron decays 
using events produced in \ppbar\ collisions at $\sqrt{s}=1.8$ TeV 
and collected by the Collider Detector at Fermilab.
The decays we considered were
$\bd \to \kst(\to K^-\pi^+)\gamma$, 
$\bs \to \phi(\to K^+K^-)\gamma$, 
$\lamb \to \Lambda(\to p\pi^-)\gamma$, 
and their charge conjugates.
Two independent methods to identify photons from such 
decays were employed. In the first method, 
the photon was detected in the electromagnetic calorimeter.
In the second method, the photon was identified by an electron-positron pair
produced through the external photon conversion before
the tracking detector volume.
By combining the two methods we obtain upper limits on the branching fractions
for the $\overline{B}_d^0$, $\overline{B}_s^0$, and $\Lambda_b^0$ radiative 
decays, which, at the 95\% confidence level, are found to be 
${\cal B}(\overline{B}_d^0\rightarrow\overline{K}^{*0}\gamma) < 1.4\times10^{-4}$, 
${\cal B}(\overline{B}_s^0\rightarrow\phi\gamma)  < 1.6\times10^{-4}$, 
and ${\cal B}(\Lambda_b^0\rightarrow\Lambda\gamma) < 1.9\times10^{-3}$.
\end{abstract}

%
%
\clearpage

\section{Introduction}

Flavor-changing neutral currents (FCNC's)
are suppressed in the Standard Model (SM) 
by the Glashow-Iliopoulos-Maiani
mechanism~\cite{GIM:1970}, 
and such transitions can only result from higher order 
processes.
The  ``penguin'' process is one such example, where an effective FCNC
$b \to s$ or $b \to d$ transition proceeds through the emission 
and reabsorption of a virtual $W$ boson.
A photon, gluon, or $Z$ boson
is emitted from the quark or the $W$ in the loop,
with the presence of a photon signaling an
``electromagnetic'' penguin process (see Figure~\ref{Fig:PengDiagram}).

It is expected in the SM that the top quark dominates 
in the fermion part of the loop of the diagram.
The existence of non-SM heavy charged particles, however, 
could affect the branching fraction for this decay.
In addition, direct $CP$-violating effects could be enhanced
by processes beyond the Standard Model.
Therefore,
measurements of radiative $b$ hadron decays, 
constitute low energy probes for 
physics beyond the SM~\cite{London:1997}.
Within the SM framework, radiative $b \to s$ decays are sensitive to the 
magnitude of the Cabibbo-Kobayashi-Maskawa (CKM) matrix~\cite{CKM} element 
$|V_{ts}|$, while radiative $b \to d$ decays are sensitive to $|V_{td}|$.
Ratios of branching fractions involving $b \to d \gamma$ and $b \to s \gamma$ 
decays can thus be used to measure the ratio $(|V_{td}/|V_{ts}|)$.
This ratio determines the length of one side of the unitarity triangle, 
and may explain the source of CP violation in the SM~\cite{Caso:1998tx}.

The branching fraction for the exclusive radiative decay 
$\Btogkst$ was first measured by CLEO to be
$(4.55 ^{+0.72}_{-0.68} \pm 0.34)\times 10^{-5}$~\cite{Coan:2000kh}.
The most precise measurements of the branching fraction
$\Br(\Btogkst)$ are 
$(4.23 \pm 0.40 \pm 0.22) \times 10^{-5}$
by the BABAR collaboration~\cite{Convery:2001pc} and
$(4.96 \pm 0.67 \pm 0.45) \times 10^{-5}$
by the BELLE collaboration~\cite{Ushiroda:2001sb}.
Both collaborations have also measured the branching fraction
$\Br(\Butogkst)$; with 
$(3.83 \pm 0.62 \pm 0.22) \times 10^{-5}$ 
obtained by BABAR~\cite{Convery:2001pc}, 
and $(3.89 \pm 0.93 \pm 0.41) \times 10^{-5}$ 
obtained by BELLE~\cite{Ushiroda:2001sb}. BELLE has also reported 
$\Br(B \to \rho \gamma)/\Br(B \to K^{*} \gamma) < 0.19$ 
at 90\% confidence level (CL)~\cite{Ushiroda:2001sb}.
The branching fraction 
for the inclusive radiative decays 
$B \to X_{s} \gamma$, where $X_{s}$ represents a collection of hadrons 
containing strange quarks, was also measured by 
CLEO to be
$(3.15 \pm 0.35 \pm 0.32 \pm 0.26)\times 10^{-4}$~\cite{Ahmed:1999fh},
where the first uncertainty is statistical, the second is systematic,
and the third is for model dependence.
  The studies of the heavier $b$-hadron decays such as \Bs\ and $\Lambda_b$, 
which are not produced at the $\Upsilon(4S)$, must be done at the
higher energy machines, such as the Tevatron.
No exclusive radiative decays of \Bs\ nor \Lamb\ have been observed to date.
From a search for \Btogphi\ decays,
the DELPHI collaboration obtained
$\Br(\Btogphi) < 7.0 \times 10^{-4}$ 
at 90\% CL~\cite{DELPHI:1996bsg_bs}.

  Even though calculations for the exclusive decay rates have higher
theoretical uncertainties compared to inclusive decay rates, ratios of 
exclusive $b \to d \gamma$ and $b \to s \gamma$ branching fractions 
can be calculated with good precision and the determination of 
$(|V_{td}/|V_{ts}|)$ is feasible with the use of exclusive 
decays~\cite{Ali:1994excl}.
This is especially useful for a hadron collider environment, where the 
experimental signature for radiative $b$ decays is much cleaner when
exclusive decays are considered.

 In this paper we report the results of a search for
$\overline{B}_d^0 \to \overline{K}^{*0}(\to K^-\pi^+)\gamma$, 
$\overline{B}_s^0 \to \phi(\to K^+K^-)\gamma$, 
and $\Lambda_b^0 \to \Lambda(\to p\pi^-)\gamma$ decays  
in events produced in \ppbar\ collisions 
at $\sqrt{s}=1.8$ TeV and recorded by the Collider Detector at Fermilab (CDF) 
during 1994--96.
Two methods to identify such decays are employed. 
 In the first method (Method I)~\cite{Kordas:2000}, 
the photon is detected in the electromagnetic calorimeter.
The trigger for this method required a minimum energy deposition in the
calorimeter and two oppositely charged tracks that were distinct from 
the calorimeter signal.
In the second method (Method II)~\cite{Tanaka:2001}, 
the photon is identified by an electron-positron pair
produced through an external photon conversion within
the tracking detector volume.
One of the conversion electrons, 
detected in the electromagnetic calorimeter,
served as a trigger for recording these events.
The $b$ hadrons are then exclusively reconstructed 
with four charged tracks.
%

%
\section{Collider Detector at Fermilab (CDF)}

Since CDF is described in detail elsewhere~\cite{Abe:1988me},
we describe here only the components relevant to this work.
In this paper we use a cylindrical coordinate system ($r, \phi, z$)
with the origin at the nominal interaction point, the $z$ axis parallel to
the nominal beam direction, $r$ the distance from the beam in the 
plane transverse to the $z$ axis, and $\phi$ the azimuthal angle.
We define $\theta$ to be the angle
with respect to the $+z$ direction and the pseudorapidity 
as $\eta\equiv-\ln[\tan(\theta/2)]$.

The tracking systems consist of a silicon vertex detector (SVX),
a vertex time projection chamber (VTX), and an open-cell multiwire
drift chamber (CTC), all immersed in a $1.4\;{\rm T}$ solenoidal magnetic
field aligned with the $z$ axis.  The SVX~\cite{CDF:1995svx}
is the innermost system, with its
four layers of single-sided silicon microstrip detectors in the
radial range of 3.0 to $7.9\;{\rm cm}$.  The active area is $51\;{\rm cm}$
long in $z$ and covers 60\% of the
$p\overline{p}$ interaction region.  The microstrips all run parallel
to the $z$ direction and therefore track charged particles in the
transverse plane. 
The SVX measures the impact parameter of tracks with respect to the beam 
line with a resolution of
$\sigma_d(p_T) = (13 + 40/p_T)\;\mu{\rm m}$, where $p_T$ is the
momentum of the track in the transverse plane in ${\rm GeV}/c$.
This precision close to the beamline helps distinguish the tracks of
$B$ decay products from those originating at the $p\overline{p}$
interaction point.

The VTX~\cite{Snider:1988az} surrounds the SVX and consists of 
28 drift modules with an
outer radius of $22\;{\rm cm}$ and $z$ coverage up to $\pm 136\;{\rm cm}$.
The VTX tracks particles in the $r-z$ plane and provides a measurement
of the actual $p\overline{p}$ interaction point along the $z$ axis with
a resolution of 1 to $2\;{\rm mm}$.  From a combination of this
information with SVX measurements, the transverse beam profile has been
measured with an accuracy of $25\;\mu{\rm m}$.

Outside the VTX lies the CTC~\cite{CDF:1988ctc},
which extends out to a radius of $138\;{\rm cm}$
and $|z|<160\;{\rm cm}$.  It contains 6156 wires arranged in 84 layers,
which are further grouped into 9 ``superlayers''.  Five of these
superlayers are made of twelve layers of wires strung parallel to the
$z$ axis (``axial superlayers'').  The remaining four superlayers of six
wires each are tilted $3^{\circ}$ in the $\phi$
direction (``stereo superlayers'').
The combination of axial and stereo measurements yields a three-dimensional
track.  Where appropriate, this track is augmented with SVX measurements
to obtain precise impact parameters.  The momentum resolution of such
tracks, often simply called ``SVX tracks,'' is
$\sigma(p_T)/p_T = [(0.0009p_T)^2+(0.0066)^2)]^{1/2}$ with $p_T$ in
units of ${\rm GeV}/c$.
With such momentum and impact parameter resolutions, along with the
narrow beam, CDF at the Tevatron is an excellent
tool for the study of $B$ physics.

The calorimetry systems of CDF lie outside the tracking systems and solenoid.
We focus on the calorimetry in the $|\eta|<1$ (``central'') region, which
is segmented into $\eta$-projective towers covering $15^\circ$ in azimuth and
$0.11$ units in $\eta$.  The inner layers of the towers,
which make up the central electromagnetic calorimeter (CEM)~\cite{CDF:1988cem}, consist of a
lead-scintillator stack 18 radiation lengths deep.  The CEM has a
resolution of $\sigma(E_T)/E_T=[(0.137)^2/E_T+(0.02)^2]^{1/2}$, where
$E_T=E\sin\theta$ and $E$ is the measured energy of the tower in GeV.
A layer of proportional strip chambers (CES) is embedded in the CEM near
shower maximum and provides measurements of shower position and profile
in azimuth and $z$~\cite{CDF:1988cem}.
The outer layers of the calorimeter tower, which  make up the central hadron 
calorimeter (CHA), consist of an iron-scintillator stack 4.5 interaction 
lengths deep and yield an energy resolution of
$\sigma(E_T)/E_T=[(0.50)^2/E_T+(0.03)^2]^{1/2}$.  In this analysis, the CHA
is used primarily to distinguish electrons and photons, which are
typically absorbed in the CEM, from hadrons, which typically deposit
most of their energy in the CHA.

A three-level trigger system is employed at CDF to select $p\overline{p}$
events of interest~\cite{CDF:1988trigger}.
The first-level trigger relevant to this analysis
selects events based on energy depositions in logical ``trigger towers''
which consist of two adjacent (in $\eta$) calorimeter towers.  The
second-level trigger forms clusters of trigger towers.  This trigger
level also incorporates a hardware track processor 
(CFT, Central Fast Tracker)~\cite{CDF:1988cft},
which searches
for tracks in the CTC using hits in the axial layers and matches those
tracks to calorimeter clusters.  The third-level trigger uses software
based on optimized offline reconstruction code to analyze the whole event.
Details of the trigger selection are given in the next section.
%

%
\section{Data}

The data used in this analysis were collected with triggers which selected
events with calorimeter signatures characteristic of electrons and
photons.  During most of the 1994-95 data-taking period (``Run IB''),
the first-level trigger selected
CEM trigger towers with minimum $E_T$ of $8\;{\rm GeV}$.  The cross
section of this trigger was $\sim 20\;\mu{\rm b}$.  

 Subsequent filtering of the surviving events was performed 
with the specialised ``penguin trigger'', which 
is a collection of requirements on all 
three final products of the penguin decay chains 
$\overline{B}_d^0\rightarrow
\overline{K}^{*0}(\rightarrow K^-\pi^+)\gamma$ and
$\overline{B}_s^0\rightarrow\phi(\rightarrow K^+K^-)\gamma$.
The innovative feature of this trigger was the use
of all the information available at the second 
trigger level to select a specific topological
configuration of the final state particles.

 The second-level trigger performed tower clustering and required
the event to
contain a cluster with $E_T > 10\;{\rm GeV}$ in the electromagnetic
section. The same cluster could include hadronic energy deposition
and the trigger required the hadronic component to be less than
12.5\% of the electromagnetic component.  
A further requirement of at least $4.5\;{\rm GeV}$ deposition in the 
CES reduced
the trigger rate by half while keeping 90\% of the electrons and photons.

The CFT track processor was then used to select topologies suggestive
of a penguin decay, with its photon and two charged hadrons.  
  No track found by the CFT was allowed to point at the
the same $\phi$ as the photon calorimeter tower (spanning $15^\circ$
in $\phi$).
Two oppositely charged tracks with $p_T>2\;{\rm GeV}/c$ 
were sought close to the photon (within two calorimeter towers)
and they were required to lie within $18^\circ$ of one another in $\phi$.
Figure~\ref{Fig:PengTrigTopology} illustrates the trigger topology.  These
track-related requirements were $\sim 35\%$ efficient for selecting
penguin events while reducing the trigger cross section to
$\sim 80\;{\rm nb}$.

When the trigger rate exceeded the limit of the data taking rate
we further reduced the trigger rate by rejecting some fraction of
the events which satisfied the trigger requirement (``prescale'').
The second-level trigger was prescaled by a factor of two whenever the
instantaneous luminosity was above $\sim 21\times 10^{30}\;{\rm cm^{-2}
s^{-1}}$.  The data loss due to the prescale, however, was minimal:
this trigger considered $(22.3\pm 0.9)\;{\rm pb}^{-1}$ out of the
$\sim 23\;{\rm pb}^{-1}$ of data available to it.

Events satisfying the second-level trigger were then 
passed to the third-level trigger for further consideration.
The photon candidate's
electromagnetic $E_T$, reevaluated with clustering software,
was required to be at least $7\;{\rm GeV}$, with an associated
hadronic energy deposition of no more than 15\% of that in the
CEM.  The profiles of energy deposition in the CEM and CES were also
required to be consistent with expectations based on test beam results for
electrons.  The track cuts applied by the second-level trigger were
confirmed at this trigger level using offline beam-constrained
tracking in the CTC.

The open points of Figure~\ref{Fig:PengTrigRate} show the penguin trigger rates
as a function of instantaneous luminosity during Run IB. 
These rates
can be compared with the total trigger rates at each trigger level, shown
by the closed points.  
From this figure we see that one out of 200 events accepted by the generic 
level-one calorimeter trigger also satisfied the second-level penguin trigger.
The third-level trigger requirements provided an additional rate reduction 
by a factor of 6.5.
Approximately 300000 events were collected during
Run IB by the penguin trigger.  
The overall trigger efficiency for penguin decays 
resulting from B mesons with $p_T > 12$ GeV/c and $|y| < 1.25$ was 
$(1.7 \pm 0.2)\%$ for $B_d \to K^{*0} \gamma$ and 
$(2.6 \pm 0.3)\%$ for $B_s \to \phi \gamma$ decays.
In the data sample collected by the penguin trigger 
in RunIB we expect around 7 $B_d \to K^{*0} \gamma$ 
and 2.6 $B_s \to \phi \gamma$ events.
This sample was further refined in the offline analysis
by selecting photon candidates in the good fiducial areas of the calorimeter,
and by requiring that full CTC track reconstruction revealed no
three-dimensional track pointing to the cluster.  
The $E_T(\gamma)$ threshold was raised to $10\;{\rm GeV}$.  
The hadronic/electromagnetic energy ratio
selection was tightened to 10\%, and requirements on shower profile
consistency were also tightened.

The trigger thresholds for the penguin trigger were lowered for the
1995-96 data-taking period (``Run IC'').  At the first trigger level,
the $E_T$ threshold was lowered to $5\;{\rm GeV}$, raising the cross
section to $\sim 30\;\mu{\rm b}$.  The second-level energy requirements were
lowered to $6\;{\rm GeV}$ in the CEM and $3\;{\rm GeV}$ in the CES
while the relative hadronic energy and track topology requirements 
were kept the same.  The trigger cross section at this level was thus 
raised to $\sim 500\;{\rm nb}$.  
The photon $E_T$ threshold was lowered to 5 GeV in the third-level trigger, 
while the other requirements were kept the same as in RunIB. 
Due to the lower photon energy requirements, the Run IC trigger acceptance 
rate was six times higher than the RunIB trigger, and the signal 
yield increased by a factor of five.
As a result of these adjustments, approximately 500000 events were collected 
from the only $(6.6\pm 0.3)\;{\rm pb}^{-1}$ of Run IC integrated luminosity.
The offline $E_T$ cut was accordingly lowered for this data to $8\;{\rm GeV}$.

A sample of electron candidates was also accumulated through Runs IB
and IC.  The trigger for this sample used the same first-level requirements
as described above, but required $E_T > 8\;{\rm GeV}$ at the second level,
along with a CFT track with $p_T > 7.5\;{\rm GeV}/c$ pointing to the
EM cluster's $\phi$ bin.  At the third trigger level, the reevaluated
thresholds were $E_T > 7.5\;{\rm GeV}$ and $p_T > 6\;{\rm GeV}/c$.
Moreover, the track's trajectory was extrapolated to the CES and compared
with the shower positions; agreements within $\pm 3\;{\rm cm}$ in the
azimuthal direction and $\pm 10\;{\rm cm}$ in $z$ were required.  These
trigger requirements were applied throughout Runs IB and IC.

The electron candidate sample serves two purposes in this analysis.
In Method~I we search for radiative decays among events selected
by the penguin trigger. The electron sample provides a reference signal,
$\overline{B}\rightarrow e^- D^0(\rightarrow K^- \pi^+) X$, 
which we compare to the yield of
radiative decay candidates.  To facilitate this comparison, the same
fiducial, $E_T$, and calorimeter requirements were applied offline to the
subsample of the electron data which was collected concurrently
with the penguin trigger; the uncertainties in the integrated luminosities
of these two data sets are thus completely correlated.  
Because this reference sample was obtained by triggering on electrons, 
a single track was required to point to the electron cluster.
Nevertheless, in order to simulate the penguin trigger requirements,
no other track was allowed to point to that $\phi$ bin.

In Method~II, where the photons are identified through their conversion
to $e^+e^-$ pairs, the search for radiative decays is performed in the
electron candidate sample itself.  In this case, the offline selection
applies fiducial, shower profile, and track-shower match requirements
in a manner similar to Method~I, but the $E_T$ threshold is lower
at $8\;{\rm GeV}$.  The minimum track $p_T$ is $6\;{\rm GeV}/c$.
The hadronic/electromagnetic energy ratio requirement is tightened
to 4\% when only one track pointed to the cluster, but is left at 10\%
in cases with more than one track associated with the cluster.  
This sample also provided the reference signal, 
$B^+\rightarrow J/\psi(\rightarrow e^+e^-) K^+$,
and thus the entire Run IB data set is used for this method.
The electron trigger accumulated $74\;{\rm pb}^{-1}$ during this period, 
amounting to approximately 3 million events satisfying the offline criteria.
%

%
\section{Method I: Photon Trigger}

In this section, we describe the search for $\overline{B}_d^0\rightarrow
\overline{K}^{*0}(\rightarrow K^-\pi^+)\gamma$ and
$\overline{B}_s^0\rightarrow\phi(\rightarrow K^+K^-)\gamma$ decays
using the penguin trigger described in the previous section.  
The sensitivity
of this method to $\Lambda_b^0\rightarrow\Lambda(\rightarrow p\pi^-)\gamma$
is strongly reduced by the trigger requirement of $p_T > 2$ GeV/$c$ for the 
pion track, because in the $\Lambda \to p\pi$ decays the proton carries 
most of the momentum of its parent and the pion is very slow.
Thus, we do not attempt to reconstruct such decays.
We derive
the branching fraction limits for the radiative $B$ decays from the ratios
between the numbers of candidate events and events of the reference signal,
$\overline{B}\rightarrow e^- D^0(\rightarrow K^- \pi^+) X$, found in the 
single electron data set.

\subsection{Radiative Decay Reconstruction}
\label{Sec:M1Sig}

We selected candidate daughters of the $\overline{K}^{*0}$ and $\phi$ mesons
from the radiative $B$ decays by asking for two oppositely charged tracks
reconstructed with the inclusion of at least three hits in the SVX.
Each track was required to have been found by the trigger system and have
$p_T>2\;{\rm GeV}/c$.  The penguin trigger topology requirements on the
tracks and the photon candidate were reinforced offline.  We then constrained
each pair of candidate tracks to intersect at a common vertex and required
the confidence level (CL) of the constrained fit to exceed 1\%.

We retained two-track combinations consistent with
$\overline{K}^{*0}\rightarrow K^-\pi^+$ by requiring
$|M(K^-\pi^+)-M_{\overline{K}^{*0}}| < 80\;{\rm MeV}/c^2$, where
$M_{\overline{K}^{*0}}$ is the world average $\overline{K}^{*0}$ mass
($896.1\;{\rm MeV}/c^2$) \cite{Caso:1998tx}.  
This window, corresponding to three times the 
natural $\overline{K}^{*0}$ width,
contained more than $85\%$ of the $\overline{K}^{*0}$ signal.
If the track pair also fell within the mass
window when the $K$ and $\pi$ mass assignments were switched, we chose the
assignment which yielded the two-track mass closer to the world average.
This approach yielded the correct assignment 88\% of the time.
For $\phi\rightarrow K^+K^-$ decays, we required
$|M(K^+K^-)-M_{\phi}|<10\;{\rm MeV}/c^2$, where $M_{\phi}$ is the world
average $\phi$ mass ($1019.4\;{\rm MeV}/c^2$) \cite{Caso:1998tx}.  
This window, corresponding to four times the natural $\phi$ width,
contained $86.5\%$ of the $\phi \to K^+ K^-$ signal.

In order to reject $K^0\rightarrow\pi^+\pi^-$ decays, we assigned pion masses
to the two tracks and required that 
$|M(\pi^+\pi^-)-M_{K^0}|>15\;{\rm MeV}/c^2$.
We thus rejected  
combinations with masses within $2\sigma$ of the world average $K^0$ mass
and retained 95.4\% of the
$\overline{K}^{*0}\rightarrow K^-\pi^+$ decays and all of the
$\phi\rightarrow K^+K^-$ decays.

The track pair was combined with the photon candidate by adding their
four-momenta.  The trajectory of the photon candidate was determined
by assuming that it originated from the $p\overline{p}$ vertex closest
in $z$ to the track pair vertex; we call this $p\overline{p}$ vertex
``primary.''  Because the lifetimes of the $\phi$ and
$\overline{K}^{*0}$ mesons are almost ten orders of magnitude smaller
than that of the $B$ meson~\cite{Caso:1998tx}, the common fitted vertex of the
two charged tracks indicated the point where the parent $B$ meson decayed.
We computed the $B$ meson's signed decay length
$L_T = \vec{V}_T\cdot\vec{p}_T/p_T$, where $\vec{V}_T$ is the
displacement in the transverse plane of the $B$ decay vertex with respect
to the primary vertex (see Figure~\ref{Fig:Bdecay}), and $\vec{p}_T$ is the
$B$ meson momentum projected on the same plane.  The proper decay length
$ct$ could then be calculated with 
$ct = L_T\cdot(M/p_T)$, where $M$ is the
reconstructed mass of the $B$ meson candidate.  The typical $ct$ resolution
was $100\;\mu{\rm m}$.  
We required $0 < ct < 3\;{\rm mm}$, which
retained 90\% of the signal while rejecting half of the fake $B$ meson 
candidates formed by tracks coming directly from the primary vertex.

We further required that the $B$ meson carry most of the momentum in its
vicinity.  We defined the isolation variable
\begin{eqnarray}
 I_B\equiv\frac{p_T(B)}{p_T(B) + \sum_{\Delta R\leq 1}p_T} \, ,
\end{eqnarray}
where the sum is over tracks consistent with originating from the
primary vertex and within
$\Delta R\equiv\sqrt{(\Delta\eta)^2+(\Delta\phi)^2} < 1$ of the $B$
candidate trajectory.  The $B$ candidate daughters were excluded from the sum.
We required $I_B > 0.7$.  Studies with reconstructed $B$ decays in data
indicate that this requirement is $\sim 95\%$ efficient in selecting real
$B$ mesons of $p_T>15\;{\rm GeV}/c$ while rejecting half of the
combinatorial background.

The mass resolution of $B$ mesons reconstructed in the above manner is
given by simulation to be $110\;{\rm MeV}/c^2$, dominated by the energy
resolution of the photon.  
We have used $D^0 \rightarrow K^- \pi^+$  and electrons from the reference 
signal $\overline{B}\rightarrow e^-D^0X$ to verify that 
the simulation closely reproduces the momentum resolution and impact parameter
resolutions of tracks, as well as the energy resolution and shower 
characteristics of electromagnetic objects.
 After the above selection criteria, there are
$\sim 400$ $\overline{K}^{*0}\gamma$ and $\sim 40$ $\phi\gamma$ events
within $220\;{\rm MeV}/c^2$ of the world average $\overline{B}_d^0$ and
$\overline{B}_s^0$ masses of 
$5279\;{\rm MeV}/c^2$ and $5369\;{\rm MeV}/c^2$ 
respectively~\cite{Caso:1998tx}.

To further improve our sensitivity to the radiative decays, we exploited the
long $B$ meson lifetime and the fact that we reconstructed all its daughters.
The long lifetime resulted in large impact parameters for the
$\overline{K}^{*0}$ and $\phi$ daughters with respect to the primary
vertex; we cut on the significance of the impact parameters in the
transverse plane, $|d/\sigma_d|$.  The impact parameter resolution was
typically $\sigma_d\sim 30\;\mu{\rm m}$.  
We also formed an ``alignment angle''
between the transverse momentum $\vec{p}_T$ and the displacement
$\vec{V}_T$ of the $B$ meson candidate (see Figure~\ref{Fig:Bdecay}):
\begin{eqnarray}
\vartheta_{{\rm align}}\equiv\cos^{-1}\left(\frac{\vec{p}_T\cdot\vec{V}_T}{
   |\vec{p}_T|\cdot|\vec{V}_T|}\right).
\end{eqnarray}
Since we fully reconstructed the $B$ meson, real mesons yielded small values
of $\vartheta_{{\rm align}}$, whereas the combinatorial background peaked away
from zero.
As a pure background sample we used events in the high mass region 
$6<M(\overline{K}^{*0}\gamma,\phi\gamma)<10\;{\rm GeV}/c^2$,
where no real $B$ mesons should be found. 
Comparing the $\theta_{align}$ distributions of the simulated signal events 
with the distribution obtained from the background sample, 
we selected signal-like events by demanding $\theta_{align} < 0.15$ rad,
for both the $\overline{B_d}$ and $\overline{B_s}$ decays. 
We subsequently found the impact parameter significance cut
which gave the highest signal-to-background efficiency ratio. 
It turned out that the best value was the one which
rejected all events in the background (high mass) region.

The optimized selection cuts for $\overline{B}_d^0$ radiative decays were
$\vartheta_{{\rm align}} < 0.15\;{\rm rad}$ and $|d/\sigma_d|~>~5$.  These
requirements were 66\% efficient in retaining
$\overline{B}_d^0\rightarrow\overline{K}^{*0}\gamma$ decays.  
For the
$\overline{B}_s^0$ decays, the narrower $\phi$ resonance, compared
to the $\overline{K}^{*0}$, resulted in a smaller number of combinatorial
background events falling within the, consequently narrower, mass window
used to select the relevant two-track pairs.
Thus, the optimized $|d/\sigma_d|$ cut for the
$\overline{B}_s^0$ was less strict at $|d/\sigma_d|>2.5$.  These optimized
requirements are 69\% efficient in retaining
$\overline{B}_s^0\rightarrow\phi\gamma$ decays.

All together, the offline analysis requirements were $14.7\%$ ($8.3\%$) 
efficient in selecting  $\overline{B}_d^0\rightarrow\overline{K}^{*0}\gamma$ 
decays which already satisfied the trigger requirements in Run IB (Run IC).
The respective efficiencies for $\overline{B}_s^0\rightarrow\phi\gamma$ decays
were $12.9\%$ in Run IB and $7.2\%$ in Run IC.
The Run IC efficiencies were lower than Run IB because the offline cut for the 
photon energy was placed at $E_T > 8\;{\rm GeV}$, instead of
the $6\;{\rm GeV}$ trigger threshold, in order to match the energy threshold 
of the electrons used to reconstruct the reference channel.

Figure~\ref{Fig:Bd_Bs_allcuts_Peng} shows the invariant mass 
distributions of the three-body
combinations surviving all the selection criteria.  The
$\pm 220\;{\rm MeV}/c^2$ signal region around the world average $B$
mass is double-hatched in the figure, and the sideband regions,
$3.9<M(\overline{K}^{*0}\gamma)<4.9$ and
$5.7<M(\overline{K}^{*0}\gamma)<6.7\;{\rm GeV}/c^2$, are single-hatched.
One $\overline{B}_d^0\rightarrow\overline{K}^{*0}\gamma$ candidate,
from the Run IC sample, remains in the signal region, while five populate
the sidebands.  The expected background in the signal region, assuming
a uniform distribution interpolated between the sidebands, is
$N_{{\rm bg}} = 1.1\pm 0.5$ events.
There are 2 events just outside the signal window. 
However, the probability of them being signal is small. 

In the $\overline{B}_s^0\rightarrow\phi\gamma$ case, no candidates
survive the selection cuts.  Since there are also no events in the
$\overline{B}_s^0$ sidebands, in the signal
region we expect $N_{{\rm bg}}<0.54$ events with 
$90\%$ confidence~\cite{Caso:1998tx},
assuming a uniform distribution interpolated between the sidebands.

\subsection{Reference Signal Reconstruction}

We reconstructed our reference sample of 
$\overline{B}\rightarrow e^- D^0(\rightarrow K^- \pi^+) X$
decays, 
by adding the four-momenta of the 
two tracks and the electron candidate.  
For $e+D^0$ combinations from $B$ decays, we expected the kaon
from the $D^0$ to have the same charge as the electron.  The mass assignment
of the pion and kaon masses to the two tracks was thus uniquely determined.

We retained $\overline{B}\rightarrow e^- D^0(\rightarrow K^- \pi^+) X$ 
candidates with a $p_T(B)$ 
distribution similar to that of the radiative decay candidates
by requiring $p_T(eK\pi)>15\;{\rm GeV}/c$ in Run IB.  For
Run IC, this threshold was lowered to $13.5\;{\rm GeV}/c$ to accommodate the
lower photon threshold.  We also required that the mass of the three-body
combination $M(eK\pi)$ be less than $5\;{\rm GeV}/c^2$.  
Finally, we applied
the same $0 < ct < 3\;{\rm mm}$ and $I_B > 0.7$ requirements as on the 
radiative decay candidates.  These semileptonic decays, however, were not
fully reconstructed, and we used the combined momentum of the $e+D^0$ system
for the (pseudo-proper) lifetime calculation.  In addition, rather than
extrapolating the
$D^0$ decay vertex to the trigger electron track in order to locate the
$B$ decay vertex, we simply used the $D^0$ decay vertex for the calculation
of $ct$ to avoid additional systematic uncertainties due to the further
vertex reconstruction.

We then required $|d/\sigma_d|>3$ for the kaon and pion tracks from the
$D^0\rightarrow K^-\pi^+$ decay.  
Since $\overline{B}\rightarrow e^-D^0X$ decays are not fully reconstructed, 
we do not make a $\vartheta_{{\rm align}}$ cut.  
The invariant masses of the selected $K^-\pi^+$ combinations from
$\overline{B}\rightarrow e^-D^0X$ candidates are shown in 
Figure~\ref{Fig:d0_all_cuts}.
The $K^-\pi^+$ combinations with the wrong charge correlation with the
electron are also shown.  We estimated the number of
$\overline{B}\rightarrow e^-D^0X$ candidates by fitting the data with a
Gaussian signal and a linear background and we found $40.7\pm 7.3$ events
in Run IB and $27.4\pm 6.2$ events in Run IC.

\subsection{Efficiencies}
\label{mi:eff}

In method I we infer the radiative decay branching fraction
from a measurement of its ratio with the known $\Br(\Btoedx)$.
The $b$-quark production cross section cancels in the ratio, 
while the effect of systematic uncertainties is reduced.
We write for \Btogkst\ ,
\BE
\label{Eq:Bd_peng_Brel}
\nonumber
\Br(\Btogkst) =
\Br(\Btoedx) \times
\frac{N_{\kstp\gamma}}{N_{\edz}}\\
\nonumber
\times \left[
\frac { f_d }{f_u + f_d}
\frac {\Br(\Ksttokpi)}{\Br(\Dtokpi)}
\frac {\epsilon_{\kstp\gamma}}{ \epsilon_{\edz} }
\frac {L_{\rm peng} }{ L_{eX}}
\right]^{-1},\\
\EE
and for \Btogphi\ , 
\BE
\label{Eq:Bs_peng_Brel}
\nonumber
\Br(\Btogphi) =
\Br(\Btoedx) \times
\frac{N_{\phi\gamma}}{N_{\edz}}\\
\times \left[
\frac { f_s }{f_u + f_d}
\frac {\Br(\Phitokk)}{\Br(\Dtokpi)}
\frac {\epsilon_{\phi\gamma}}{ \epsilon_{\edz} }
\frac {L_{\rm peng} }{ L_{eX}}
\right]^{-1},
\EE
where 
$\left. N_{\kstp\gamma,\phi\gamma} \right/ N_{\edz}$ is the ratio of the 
observed number of events of the radiative decays and \Btoedx,
$\left. \epsilon_{\kstp\gamma,\phi\gamma} \right/ 
\epsilon_{\edz}$ is the ratio of the efficiencies, and
$\left. L_{\rm peng} \right/ L_{eX}$ is the ratio of 
the integrated luminosities of the penguin and the inclusive electron data 
samples.
We assume that the composition of \Btoedx\ candidates 
is only $B_u^-$ and $\overline{B}_d^0$, and thus the ratios of the
fragmentation fractions are $f_{d,s}/(f_u+f_d)$, neglecting the small
contributions from other $b$ hadrons such as $\overline{B}_s^0$ and
$\Lambda_b^0$ to the denominator.  We note that the contribution of
the $\overline{B}_s^0$ through the
$\overline{B}_s^0\rightarrow e^-\overline{\nu}_eD_s^{**}\rightarrow e^-D^0X$
decay is estimated to be less than 3\% in the $e+D^0$ sample.
The branching fractions~\cite{Caso:1998tx} and
fragmentation fractions~\cite{Affolder:2000iq}
used in this analysis are listed in Table~\ref{Table:BRrel_BC_Peng}.

Since we use electron trigger data collected concurrently with 
the penguin trigger data,
the integrated luminosities
of \ppbar\ collisions are the same for the two data sets. 
The effective integrated luminosities of each data set, however,
are different for the two due to the different 
prescale factors. 
The true integrated luminosities for 
the penguin and electron data set are 22.3 pb$^{-1}$ 
and 16.2 pb$^{-1}$, respectively, in Run \ib,
and 6.6 pb$^{-1}$ and 4.2 pb$^{-1}$ in \ic.
We assume that all the uncertainties cancel in the ratio.

The efficiency ratios were evaluated using a combination of simulation and
data.  We employed a Monte Carlo simulation of events with a single $b$ quark
to calculate the efficiencies of the kinematic and topological requirements
imposed on the data.  In this simulation, the $b$ quarks were generated with
a rapidity and momentum distribution based on a next-to-leading order QCD
calculation~\cite{Nason:1989zy} that used the MRSD0 parton distribution
functions~\cite{Martin:1993as} and a renormalization scale of
$\mu=\mu_0\equiv\sqrt{m_b^2+p_T^2}$, where $m_b=4.75\;{\rm GeV}/c^2$ is the
mass of the $b$ quark and $p_T$ is its transverse momentum.  These $b$
quarks were subsequently hadronized into $B$ mesons using the Peterson
fragmentation function~\cite{Peterson:1983ak} with a fragmentation parameter
$\epsilon_b=0.006$.  The resulting $B$ mesons were then decayed through
the channel of interest using the QQ Monte Carlo program~\cite{Avery:1985} to
model the phase space, helicity, and angular distributions of the decay
products.  

For the reference channel, we generated different samples for
each of the contributing decay chains:
$\overline{B}\rightarrow e^-\overline{\nu}_eD^0$;
$\overline{B}\rightarrow e^-\overline{\nu}_eD^*(\rightarrow D^0X)$;
$\overline{B}\rightarrow e^-\overline{\nu}_eD^{**}(\rightarrow D^0X)$;
and $\overline{B}\rightarrow e^-\overline{\nu}_e(Dn\pi)_{{\rm nr}}$ followed
by $(Dn\pi)_{{\rm nr}}\rightarrow D^0X$, where $(Dn\pi)_{{\rm nr}}$ indicates
a $D$ meson produced in non-resonant association with extra pions.  We
then mixed these semileptonic samples according to their relative abundances
and selection efficiencies to create a representative
$\overline{B}\rightarrow e^-D^0X$ sample.  We fed these events through
the detector and trigger simulations to obtain the efficiencies.  We also
used this simulation to calculate the relative effects of the photon/electron
trigger cuts, the offline quality cuts, and the track reconstruction in
the SVX. We considered simulated SVX track reconstruction since the SVX 
simulation incorporated the same hit efficiencies
and pattern recognition as the data.

Second-level trigger efficiencies were studied using data.  The efficiency of
the CES energy requirement was parameterized as a function of electron or
photon $E_T$ by analyzing electrons in a very pure sample derived from
photon conversions.  Applying this parameterization to the Monte Carlo
samples, we find all the efficiencies to be around 95\%.  The efficiency
ratios are therefore near unity, and the 2\% uncertainty in the ratio
is included in the systematic uncertainty.

The efficiency of the CFT trigger requirements for kaons and pions
was determined as a function of track $p_T$. 
We found the CFT is 50\% efficient at $1.9\;{\rm GeV}/c$ and 90\% efficient
at $2.4\;{\rm GeV}/c$.  
The efficiency function of the CFT trigger requirements for the electron in 
the 
reference signal was determined using a heavily prescaled electron data set 
with a lower energy threshold and no CFT requirement; 50\% efficiency is reached at
$6.0\;{\rm GeV}/c$ and 90\% at $10.0\;{\rm GeV}/c$.  The plateau
efficiency is $0.915\pm 0.010 $.  These efficiency parameterizations were
applied to the Monte Carlo samples to study the effect on the ratios of
efficiencies.

The offline CTC tracking efficiencies for kaons and pions were estimated
by embedding Monte Carlo-generated tracks into real
$J/\psi\rightarrow\mu^+\mu^-$ events~\cite{Abe:1998andreas}.  The efficiency
rises with $p_T$ in the range $200<p_T<400\;{\rm MeV}/c$, and plateaus at
a value which depends on the instantaneous luminosity and the
charge of the track.  
The integrated efficiency for tracks with $p_T>400\;{\rm MeV}/c$ 
is $0.96\pm 0.02$.  Again, we applied the efficiency parameterization
to Monte Carlo samples of the decays of interest.  
For $\overline{K}^{*0}\rightarrow K^-\pi^+$ and 
$\phi \rightarrow K^+ K^-$ decays
with the requirement $p_T>2\;{\rm GeV}/c$ for the kaons and pions,
the efficiency of offline CTC tracking was found to be $0.94\pm 0.04$.  
The corresponding efficiency
for the $K^- \pi^+$ combinations from the $D^0$ decays is $\sim 1\%$ lower due
to the lower $p_T$ of the tracks.  The uncertainties in these efficiencies
are dominated by the instantaneous luminosity dependence of the tracking
efficiency and thus cancel in the efficiency ratio.
The offline tracking efficiency for the trigger electron in the
reference signal was estimated using an independent electron data sample to
be $0.99\pm 0.01$.  We therefore estimate the ratio of tracking efficiencies
for both $K^{*0}\gamma$ and $\phi\gamma$, relative to the reference signal,
to be $1.02\pm 0.02$.

The effect of the isolation requirements for the trigger photon or electron, as
well as the $I_B>0.7$ cut for the $B$ meson, depends strongly on the
environment of the $B$ decay (e.g., $b$ fragmentation, or multiple
$p\overline{p}$ interactions).  We expect similar environments around the $B$
mesons in the reference and radiative decay processes and
consequently the efficiencies are nearly equal.  Small differences can be
expected due to the extra particles produced in
$\overline{B}\rightarrow e^-D^0X$ decays and because the reference
signal contains $B^-$ mesons along with $\overline{B}^0$.
We simulated the full $p\overline{p}\rightarrow b\overline{b}$ environment
using the PYTHIA Monte Carlo generator, tuned to match the underlying charged
particle distributions in $\overline{B}\rightarrow\ell^-D^0X$
data~\cite{Abe:1999sq}.
We fed these events through the detector and trigger
simulations and found that the isolation efficiencies are
somewhat higher for the radiative decay channels than for the reference
signal; the ratio is $1.04\pm 0.02$ for
$\overline{B}_d^0\rightarrow\overline{K}^{*0}\gamma$ and $1.06\pm 0.02$ for
$\overline{B}_s^0\rightarrow\phi\gamma$.

Taking all the efficiencies into account, we find that the efficiency
ratios between the radiative decays and the reference channel are
$\epsilon_{K^{*0}\gamma}/\epsilon_{eD^0}=2.7$ in Run IB and 2.0 in Run IC.
In the $\overline{B}_s^0\rightarrow\phi\gamma$ case, we find these ratios
to be 3.5 in Run IB and 2.5 in Run IC.

Table~\ref{Table:BRrel_BC_Peng} summarizes the elements of the 
branching fraction
calculation for each of the decay modes investigated here.  The table
also shows the ``single event sensitivity'' $S$ for the two penguin decay 
modes. $S$ is defined here as
\begin{equation}
  S = \frac{{\rm branching\;fraction}}{{\rm observed\;number\;of\;events}}
\label{Eq:Sensitivity}
\end{equation}
and can be rewritten with the known quantities by 
using Eqs.~\ref{Eq:Bd_peng_Brel} and \ref{Eq:Bs_peng_Brel}.  
This quantity represents the branching fraction which
would result in an average of one event being observed in this analysis.
The difference in the sensitivities between the 
$\overline{B}_d^0\rightarrow\overline{K}^{*0}\gamma$ and
$\overline{B}_s^0\rightarrow\phi\gamma$ decay modes is
dominated by the difference of the $b$ quark hadronization fractions.

\subsection{Systematic Uncertainties}
\label{Sec:syst}

Table~\ref{Table:syst_unc_BC_Peng}  lists the sources of systematic uncertainty
considered in this analysis.  The largest contribution to the total
is the uncertainty on the yield of 
$\overline{B}\rightarrow e^- D^0(\rightarrow K^- \pi^+) X$
decays, which is 18\% in Run IB and 23\% in Run IC.  The second largest
contribution arises from the uncertainty in the measurement of
$f_s/(f_u+f_d)$~\cite{Affolder:2000iq}, followed by the uncertainty in the
product of branching fractions ${\cal B}(\overline{B}\rightarrow e^-D^0X)\cdot
{\cal B}(D^0\rightarrow K^-\pi^+)$. 

The last significant contribution to the systematic uncertainty comes from
the fraction of the time when the $D^0$ meson from a $\overline{B}$ decay is
not an immediate daughter of the $\overline{B}$ meson but is instead a
decay product of an intermediate excited $D$ state.  Depending on how
far down the decay chain of the $\overline{B}$ meson the $D^0$ appears,
the kinematics of the resulting kaon and pion, and hence the reconstruction
efficiencies, are different.  In the Monte Carlo simulation used to
determine the efficiency ratios, the nominal fractions of $D^0$ mesons coming
from $D^{**}$ mesons and $(Dn\pi)_{{\rm nr}}$ states ($f^{**}$),
from $D^*$ mesons ($f^*$), and directly from the $\overline{B}$
meson ($f$) were $f^{**}:f^*:f = 0.35:0.53:0.12$~\cite{Caso:1998tx}.  
These fractions
were varied to $0.24:0.62:0.14$ and $0.47:0.43:0.10$.  We observed a 12\%
variation in the efficiency in Run IB and 11\% in Run IC.  We take these
variations as the systematic uncertainties in the efficiency ratios.

The rest of the systematic uncertainty contributions have little effect
on the total, which is about 30\%. 
For instance, the Monte Carlo efficiency estimates depend on their input
distributions, such as the $p_T$ distribution of the incident particles.
We re-weight the Monte Carlo $p_T(B)$ distribution which is used as the 
simulation input by the ratio of the
measured $B$ production cross section~\cite{Abe:1995xsec} to the theoretical
prediction.  Even though the
efficiencies for individual channels vary by as much as $\sim 20\%$, the
ratios of efficiencies do not change by more than 5\%.

Another relatively small effect is the uncertainty in the difference in
trigger efficiencies for photons and electrons.  The difference resulting
from the different $E_T$ spectra of the photons and electrons
is accounted for in the Monte Carlo calculation; moreover, we confirm
that the detector simulation indeed reproduces the characteristics of the
electromagnetic shower profile using $\overline{B}\rightarrow e^-D^0X$
decays in data.  We nevertheless assign an uncertainty due to the $E_T$
differences between the reference channel electron 
and the radiative decay photon
to allow for uncertainties in the simulation of the electromagnetic energy
clustering at the trigger level. 
We study the effect of varying the relative efficiency by re-weighting 
the photon and electron $E_T$ distribution in the lowest 10 GeV,
away from the efficiency plateau, by as much as a factor of two 
(e.g., the weight is applied for $10 < E_T < 20$ GeV in Run IB).
No weighting is applied for energies in the plateau region.
Such a modification of the threshold induces a change in the individual 
event rates by as much as 50\%, but the ratio varies by only $\sim8\%$, 
which we take as the systematic uncertainty.

The efficiency of the CES trigger requirement itself is measured with
an uncertainty of $\sim 1.5\%$.  Assuming that the efficiency for
electrons is uncorrelated with that of the radiative decay photons, we
obtain a conservative 2\% systematic uncertainty from this source.

The CFT efficiency was measured with an uncertainty 
of $\sim 1.5\%$ for kaons and pions, and 1\% for electrons.  
Due to the spatial proximity
of the two tracks in the radiative decays, we consider their efficiencies
to be 100\% correlated and thus assign a 3\% uncertainty for the
efficiency ratio.  Another 2\% uncertainty comes from the CTC tracking
efficiency, 2\% from the differences in the isolation efficiencies, and
2\% from the finite size of the Monte Carlo samples used to calculate
the efficiency ratios.

The uncertainties listed above were combined in quadrature to obtain
the total systematic uncertainties on the branching fractions of the
radiative decays.  As shown in Table~\ref{Table:syst_unc_BC_Peng}, 
the total is $\sim 30\%$ 
for $\overline{B}_d^0$ and slightly higher for $\overline{B}_s^0$.

We combine the Run IB and IC systematic uncertainties by assuming that the
uncertainties due to the statistics of the $e+D^0$ candidates and Monte
Carlo samples are uncorrelated, and any other sources are fully correlated.
The uncorrelated systematic uncertainties are added in quadrature, while
the fully correlated ones are simply added.  The total systematic
uncertainties are 25\% for $\overline{B}_d^0$ and 31\% for
$\overline{B}_s^0$ radiative decays.

\subsection{Results}
\label{Sec:MethodIResults}

Since we observe no significant signal for either $\overline{B}_d^0$ or
$\overline{B}_s^0$ radiative decays, we set upper limits for their
branching fractions.  
We use a conservative procedure which ignores possible background 
contributions to the observed event yields.

First, we calculate an upper limit on the mean number of radiative decays
$N_{{\rm CL}}$ at a given CL, including the total systematic uncertainty
$\sigma_{{\rm syst}}$, by numerically solving the following equation:
\begin{equation}
  1 - {\rm CL} = \sum_{n=0}^{N_{{\rm obs}}}
  {\cal P}_{N_{{\rm CL}},\sigma_{{\rm syst}}}(n),
\end{equation}
where $N_{{\rm obs}}$ is the number of candidates observed, and
${\cal P}_{\mu,\sigma}(n)$ is defined with the Poisson distribution
$P_\mu(n)$ and the Gaussian distribution $G_{\mu,\sigma}(x)$ as follows:
\begin{equation}
  {\cal P}_{\mu,\sigma}(n) = \int_0^\infty P_x(n)G_{\mu,\sigma}(x)dx.
\label{Eq:PG}
\end{equation}

With one \Btogkst\ candidate 
observed in the entire data sample and a 25\% uncertainty,
the upper limit on the mean number of radiative decays
is 4.3 (5.5) at 90\% (95\%) CL.
This result, with a single event sensitivity (Equation \ref{Eq:Sensitivity}) 
of $3.8 \times 10^{-5}$, yields upper limits on the
branching fraction $\Br(\Btogkst)$
of $1.6 \times 10^{-4}$ at 90\% CL and 
$2.1 \times 10^{-4}$ at 95\% CL.
With no \Btogphi\ candidates and a total uncertainty of 31\%, 
we expect less than 2.6 (3.6) events on average at 90\% (95\%) CL.
With a single event sensitivity of $9.3 \times 10^{-5}$,
we thus obtain $\Br(\Btogphi) < 2.5 \times 10^{-4}$ at 90\% CL
and $< 3.3 \times 10^{-4}$ at 95\% CL.
%

%
\section{Method II: Photon Conversion}

In this section, we describe the search for
$\overline{B}_d^0\rightarrow\overline{K}^{*0}(\rightarrow K^-\pi^+)\gamma$,
$\overline{B}_s^0\rightarrow\phi(\rightarrow K^+K^-)\gamma$, and
$\Lambda_b^0\rightarrow\Lambda(\rightarrow p\pi^-)\gamma$ decays in which
the photon is identified by an electron-positron pair produced through
photon conversion before reaching the CTC volume.  A conversion daughter
with $E_T>8\;{\rm GeV}$ served as the trigger; the same inclusive electron
trigger was used for the $e+D^0$ sample in Method~I.

Though the typical photon conversion probability was 6\% for CDF in this data,
this analysis benefits from the fact that 
we can utilize all of the Run IB 
data, which corresponds to an integrated luminosity of $74\;{\rm pb}^{-1}$, or
three times more than that collected with the penguin trigger, and 
that there was no requirement of any additional tracks at the trigger level.  
This fact allowed us to apply, in the offline selection, 
a $p_T$ threshold as low as $0.5\;{\rm GeV}/c$ to the hadron tracks coming 
from the $b$ hadron decays
instead of the $2\;{\rm GeV}/c$ cut used in Method~I.  
This lower threshold essentially doubles the efficiency for the $B$ hadron
decay products.
Moreover, in the relatively low energy region of our interest where the 
tracking has better
resolution than the calorimetry, reconstructing $b$ hadron masses from the
momenta measured by the tracking detectors has the advantage of good mass
resolution.  This is typically $45\;{\rm MeV}/c^2$ for the
reconstructed $B$ mesons and is dominated by the momentum resolution of the
trigger electron.

We derive the branching fractions for the radiative $b$ hadron decays from the
ratios between the numbers of such decays and
$B_u^-\rightarrow J/\psi(\rightarrow e^+e^-)K^-$ decays found in the same
data set.  The uncertainties in the $b$ quark production cross section and
on the integrated luminosity thus cancel, as well as most of the uncertainties
on the detection efficiency.  It would have been preferable to use
$\overline{B}_d^0\rightarrow J/\psi K^{*0}$,
$\overline{B}_s^0\rightarrow J/\psi\phi$, and
$\Lambda_b^0\rightarrow J/\psi\Lambda$ decays instead of
$B_u^-\rightarrow J/\psi K^-$, since they arise from the same production
mechanisms as the corresponding radiative decays and are topologically
more similar. However our samples of those final states are too small to be
useful as normalization.

\subsection{Radiative Decay Reconstruction}

Reconstruction of the radiative decays began with identification of a photon conversion.
A photon conversion candidate was formed by the 
electron candidate and an oppositely charged track with $p_T>0.5$ GeV/$c$.
A fit was made which constrains the two tracks
to originate from a common vertex and be parallel to each other at the vertex.
The CL of the fit was required to be greater than 0.1\%.
The background due to misidentified electrons and 
combinatorial backgrounds is small ($< 1\%$)
among the photon conversion candidates with a vertex outside the beam pipe.
The candidates that have their conversion points inside the beam pipe
are dominated by real electron-positron pairs from 
Dalitz $\pi^0$ and $\eta$ decays.
We required the transverse distance of the conversion point from 
the nominal beamline to be less than 30 cm in order to ensure that it is in
the well known materials before the CTC, and
to be greater than 3 cm in order to reject backgrounds from Dalitz decays. 
We obtained $\sim 850000$ photon conversion candidates in the Run IB data.
Figures~\ref{Fig:ConvCandxy} and \ref{Fig:ConvCandrz} show, 
for all transverse distances, the reconstructed conversion vertex density
in the $x-y$ plane and $r-z$ plane.
The fine structure of the CDF tracking detectors such as 
the SVX ($r\sim 5$ cm), the VTX ($r \sim 15$ cm), and the CTC ($r > 30$ cm) 
can be clearly resolved.
The detailed study of the CDF material distribution using $\sim 200000$
conversion candidates in 1992-1993 data is described in~\cite{Affolder:2000bp}.

For each photon conversion candidate in an event,
we searched for \Btogkst\ and \Btogphi\ decays.
A \Bd\ candidate was formed by the photon conversion candidate and 
a pair of oppositely charged tracks. The two ``meson tracks'' were required 
to be reconstructed in the SVX with hits in at least 3 layers.
In addition, the transverse momenta had to exceed 0.5 GeV/$c$
for each track and 2 GeV/$c$ for the two-track system.
A fit was performed with the following topological constraints:
(1) the meson tracks originate from a common vertex; 
(2) the photon conversion candidate points back to the meson decay vertex;
and (3) the four-track system points back to the primary vertex, which was
defined to be the \ppbar\ collision point 
nearest in $z$ to the trigger electron track's closest approach to the 
beamline.
We required the CL of the fit to be greater than 0.1\%.
The \Bd\  candidate was then accepted 
if the reconstructed \Kst\  mass was 
within $\pm 80$ MeV/$c^2$ of the world average value. 
Both $K^+\pi^-$ and $\pi^+K^-$ mass assignments were considered for
the \Kst\ candidate, and the assignment giving a value
closer to the world average was chosen.
We also required that the pseudorapidity of the $B$ candidate $|\eta_B|$ be less than 1.
Finally, we selected candidates with lifetime $ct> 100$ $\mu$m 
and $I_B > 0.7$ (See Section \ref{Sec:M1Sig}).

The selection of $\overline{B}_s^0$
candidate proceeded on similar lines, except both tracks were assigned
kaon masses and the mass window was $\pm 10\;{\rm MeV}/c^2$ around
the world average.

At this point, there were 15 $\overline{K}^{*0}\gamma$ and one $\phi\gamma$
events within $100\;{\rm MeV}/c^2$ of the corresponding world average $B$
masses.  As previously noted, the mass resolution of the reconstructed
$B$ mesons is about $45\;{\rm MeV}/c^2$.  We refined this selection by
tightening the $p_T$ cut on the two-track system and by applying impact
parameter significance cuts to the individual meson tracks.  The thresholds
were optimized by maximizing $\epsilon_{{\rm sig}}/\sqrt{\epsilon_{{\rm bg}}}$,
where $\epsilon_{{\rm sig}}$ and $\epsilon_{{\rm bg}}$ are the efficiencies
for the signal and background events found in the $\pm 100\;{\rm MeV}/c^2$
window around the $B$ masses.  The signal efficiency was obtained from
Monte Carlo calculations similar to that of Method~I (see 
Section~\ref{mi:eff}),
while $\epsilon_{{\rm bg}}$ was estimated by interpolating the observed
yields in the mass sidebands, defined to extend from 200 to
$1200\;{\rm MeV}/c^2$ above and below the average mass, through the
signal region.  For the $\overline{B}_d^0$ channel, the optimized selection
cuts were $p_T(K\pi)>2.75\;{\rm GeV}/c$ and $|d/\sigma_d|>4.5$ for both
meson tracks.  
 Figure~\ref{Fig:KstarGammaConv}(top) shows the $\overline{K}^{*0}\gamma$ mass
distribution after these cuts.  Any further cuts, for example on the
proper decay length, did not improve
$\epsilon_{{\rm sig}}/\sqrt{\epsilon_{{\rm bg}}}$.  One candidate remained
in the $\overline{B}_d^0$ signal region; the expected background is
$0.6\pm 0.3$ events.

For the $\overline{B}_s^0$ channel, the optimized selection cuts were
$p_T(KK)>2.25\;{\rm GeV}/c$ and $|d/\sigma_d|>3.0$.  The resulting
invariant mass distribution is shown in Figure~\ref{Fig:PhiGammaConv}(top).  No candidates
were found in the signal region, where we expected a background of
$0.1\pm 0.1$ events.

The decay $\Lambda_b^0\rightarrow\Lambda(\rightarrow p\pi^-)\gamma$
is topologically distinct from the meson decays.  Since the $\Lambda$ has
a long lifetime, with $ct\sim 8\;{\rm cm}$, it decays outside
the SVX fiducial volume $\sim 85\%$ of the time, and thus only 15\% of
the $\Lambda$ decays are expected to have associated SVX tracks.  We
therefore first reconstructed $\Lambda$'s without using SVX information.
The higher-$p_T$ track of the track pair was assumed to be the proton,
and was required to have $p_T>1.5\;{\rm GeV}/c$ while the pion had to have
$p_T>0.4\;{\rm GeV}/c$.  The energy loss $dE/dx$ for both tracks had to be
consistent with expectations.  A vertex-constrained fit of the track pair was
accepted if its CL exceeds 0.1\%.  Photon conversions, a major source of
background for $\Lambda\rightarrow p\pi^-$ decays, were rejected here by
eliminating those track pairs which could be fit with the conversion 
hypothesis.
 Finally, the track pair was accepted as a ``CTC-$\Lambda$'' candidate if the
distance of the decay vertex from the nominal beamline exceeded $1\;{\rm cm}$.

If both the proton and pion tracks had at least two SVX hits,
the vertex-constrained fit was redone using the SVX information.
Again, the CL of the fit was required to be greater than 0.1\%.  
We also required
the SVX layer hit pattern to be consistent with the expectation from
the reconstructed $\Lambda$ decay.  For example, if the $\Lambda$ decay
vertex was between the second and third of the four SVX layers, 
we required that the tracks have exactly two hits in the outermost layers.  
About 10\% of the ``CTC-$\Lambda$'' candidates satisfied the above 
requirements and were thus reclassified as ``SVX-$\Lambda$'' candidates.

A $\Lambda_b^0$ candidate was formed by a photon conversion and a $\Lambda$
candidate.  From the CTC-$\Lambda$ candidates, we reconstructed
``CTC-$\Lambda_b^0$'' candidates
with a constraint that both the $\Lambda$ and the photon point back to
the primary vertex.  This constraint improved the $\Lambda_b^0$ mass
resolution from $75\;{\rm MeV}/c^2$, without the constraint, to
$50\;{\rm MeV}/c^2$.  For the SVX-$\Lambda$ candidates, however, only the
photon was constrained to point back to the primary vertex, while the
$\Lambda$ trajectory was required only to point backwards to within
$2\;{\rm cm}$ in $z$ of the primary vertex.  The typical $\Lambda_b^0$ mass
for these ``SVX-$\Lambda_b^0$'' candidates is also $50\;{\rm MeV}/c^2$.  In
both cases, we required the CL of the constrained fit to exceed 0.1\%.
We then recalculated the $\Lambda$ mass given the constraints and required
that it fell within $\pm 3\;{\rm MeV}/c^2$ of the world average $\Lambda$
mass.  The typical $\Lambda$ mass resolutions are $2.5\;{\rm MeV}/c^2$
for CTC-$\Lambda_b^0$ candidates, and $1.5\;{\rm MeV}/c^2$ for
SVX-$\Lambda_b^0$.

We improved the sample purity by requiring large impact parameters,
recalculated after the constrained fit, for the proton and pion tracks.
In the SVX-$\Lambda_b^0$ case, the impact parameter resolution was good
enough to require at least $3\sigma_d$ inconsistency with the primary
vertex.  In the CTC-$\Lambda_b^0$ case, however, we noted that the proton
carries most of the momentum of its parent and required only $|d/\sigma_d|>0.5$
inconsistency. The pion from $\Lambda^0$ decay is more likely to have a large
impact parameter, so we required $|d/\sigma_d|>2$.  Finally, we selected
$\Lambda_b^0$ pseudorapidity $|\eta_{\Lambda_b^0}|<1$ and isolation
$I_B>0.7$, as before.  After these selection cuts, we found 23
CTC-$\Lambda_b^0$ and 2 SVX-$\Lambda_b^0$ candidates in the
$\pm 100\;{\rm MeV}/c^2$ window around the world average $\Lambda_b^0$
mass.

The SVX-$\Lambda_b^0$ candidates were further refined by considering the
signed impact parameter of the $\Lambda$'s.  The sign is defined as
positive when the crossing point of the $\Lambda$ and the $\Lambda_b^0$
momenta lies in the hemisphere containing the $\Lambda_b^0$, as should
be the case for real $\Lambda_b^0$ decays.  The typical resolution of
the signed impact parameter is $40\;\mu{\rm m}$. Following the same
optimization procedure as before, we find that a cut value of $70\;\mu{\rm m}$
maximizes $\epsilon_{{\rm sig}}/\sqrt{\epsilon_{{\rm bg}}}$.  No candidates
survived this cut, while the expected background is $0.1\pm 0.1$ events.

Since the CTC-$\Lambda_b^0$'s lack the improved impact parameter resolutions
of the SVX, we reinforced the kinematic requirements by requiring
the $p_T$ of the $\Lambda$ to be greater than $4\;{\rm GeV}/c$.  Two
candidates remained in the signal region, and the expected background is
$3.3\pm 0.6$ events.  Combining the CTC and SVX samples, we found two
candidates in the signal region with an expected background of
$3.4\pm 0.6$ events.  The invariant mass distribution is shown in
Figure~\ref{Fig:LambdaGammaConv} (top).

\subsection{Reference Signal Reconstruction}

The reference signal for this analysis method
consists of $B_u^-\rightarrow J/\psi(\rightarrow e^+e^-)K^-$ decays.
A $J/\psi\rightarrow e^+e^-$ candidate was formed by the electron candidate
and an oppositely charged track with $p_T > 1\;{\rm GeV}/c$.  We required
the partner track to exhibit energy loss in the CTC and deposition in
the CEM in a manner consistent with being an electron.  The two tracks
were then subject to a vertex-constrained fit, and its CL is required to be
greater than 0.1\%.  The dielectron invariant mass distribution is shown
in Figure~\ref{Fig:JpsieeCandidates}.  The ratio of signal to background $S/B$ is approximately
1/2 in the 2.8 to $3.2\;{\rm GeV}/c^2$ mass range.  The backgrounds are
mostly combinatorial, involving hadrons misidentified as the partner
electron.  The low-mass tail on the signal is due to photon
bremsstrahlung on the electron tracks.  A fit of the mass distribution
with two Gaussians and a second-order polynomial yields
$\sim 8000\; J/\psi\rightarrow e^+e^-$ events.

The $J/\psi$ candidates were then combined with a track with
$p_T>2\;{\rm GeV}/c$.  We required that all three tracks incorporate at
least 3 SVX hits.  We constrained the tracks to a common vertex
pointing back to the primary vertex and accepted the
combination if the CL of this fit exceeded 0.1\%.  We also required that
the candidate trajectory fall within the pseudorapidity range
$|\eta_B|<1$, have proper lifetime $ct>100\;\mu{\rm m}$, and
isolation $I_B>0.7$.  The resulting $M(eeK)$ mass distribution shows
the same low-mass bremsstrahlung tail as the $M(ee)$ distribution; in
order to correct for it and, at the same time, compensate for the
resolution lost because of the electron momentum uncertainty, we plot
$M(eeK)-M(ee)+M_{J/\psi}$, where $M_{J/\psi}$ is the world average
$J/\psi$ mass, instead of $M(eeK)$.  The resolution on this compensated
mass is typically $25\;{\rm MeV}/c^2$, whereas it is typically
$50\;{\rm MeV}/c^2$ for $M(eeK)$ alone.

After the above selection, we have $48\;J/\psi$ candidates with
$S/B\sim 10$ in the $\pm 100\;{\rm MeV}/c^2$ window around the world
average $B_u^-$ mass.  Further requirements, determined by the cut
optimizations on the different radiative decays, were applied to this sample
in order to achieve as much cancellation of the systematic uncertainties
as possible. 
To compare with the
$\overline{B}_d^0\rightarrow\overline{K}^{*0}\gamma$ decays shown in
Figure~\ref{Fig:KstarGammaConv}(bottom), these
requirements are $p_T(K)>2.75\;{\rm GeV}/c$ and $|d(K)/\sigma_d|>4.5$.
The signal yields were calculated by subtracting the 
backgrounds
estimated from the sidebands, which range from 200 to $300\;{\rm MeV}/c^2$
above and below the $B_u^-$ mass. 
The yield is $28.0\pm 5.8$ events.  In the
$\overline{B}_s^0\rightarrow\phi\gamma$ case shown in
Figure~\ref{Fig:PhiGammaConv}(bottom), the cuts are $p_T(K)>2.25\;{\rm GeV}/c$ and
$|d(K)/\sigma_d|>3$, yielding $35.0\pm 6.4$ events.  For the
$\Lambda_b^0\rightarrow\Lambda\gamma$ case, only the $p_T(K)>4\;{\rm GeV}/c$
cut was applied.  
The yield, shown in Figure~\ref{Fig:LambdaGammaConv}(bottom), is
$24.0\pm 5.3$ events.

\subsection{Efficiencies}

Because no significant excesses over backgrounds were observed in any of the
radiative decay modes investigated, we set upper limits on the
branching fractions.  As in Method~I, we start from the ratios between the
number of observed signal and reference decays. 
Since these decays were reconstructed
in the same data set, the $b$ quark production cross section and the
integrated luminosity of the data cancel in this ratio.  The fragmentation
fractions, branching fractions, and total reconstruction efficiencies,
on the other hand, do not cancel in principle, and their ratios must be
estimated.  We write the following relations:
\begin{eqnarray}
\label{Eq:Kstg}
\nonumber
\Br(\Btogkst)
&=&
\Br(\Btojpsik)\times
\frac
{N_{K^{*0}\gamma}}{N_{\jpsi K}}\\
&\times&
\left[
\frac{f_d}{f_u} 
\frac{ \Br(\Ksttokpi)}{\Br(\Jpsitoee)}
\frac{\epsilon_{K^{*0}\gamma}}{\epsilon_{\jpsi K}}
\right]^{-1},\\
\label{Eq:Phig}
\nonumber
\Br(\Btogphi)
&=&
\Br(\Btojpsik)\times
\frac
{N_{\phi\gamma}}{N_{\jpsi K}}\\
&\times&
\left[
\frac{f_s}{f_u} 
\frac{\Br(\Phitokk)}{\Br(\Jpsitoee)} 
\frac{\epsilon_{\phi\gamma}}{\epsilon_{\jpsi K}}
\right]^{-1},\\
\label{Eq:Lamg}
\nonumber
\Br(\Lambtoglam)
&=&
\Br(\Btojpsik)\cdot
\frac
{N_{\lam\gamma}}{N_{\jpsi K}}\\
&\times&
\left[
\frac{f_{\lamb}}{f_u} 
\frac{ \Br(\Lamtoppi)}{\Br(\Jpsitoee)}
\frac{\epsilon_{\lam\gamma}}{\epsilon_{\jpsi K}}
\right]^{-1}.
\end{eqnarray}
The branching fractions~\cite{Caso:1998tx}
and fragmentation fractions~\cite{Affolder:2000iq}
which we used are listed in Table~\ref{Table:EfficienciesConv}.
 The remainder of the calculation concerns the efficiency ratios.  The
efficiency ratios for most kinematic and geometric requirements,
including those on $E_T$, $p_T$, masses, $ct$, impact parameters,
and fit constraints, can be reliably calculated with simulation,
as in Method~I.  Likewise, the effect of the electron trigger can be
calculated by applying an efficiency curve as a function of electron
$E_T$ and $p_T$ to the Monte Carlo samples, where the curve is based
on measurements using unbiased data collected with independent triggers.
We assume that the $B$ isolation cut efficiencies cancel exactly in
the ratio, since, unlike in Method~I, the reference decay is fully
reconstructed.

The effect of the tracking efficiencies on the ratio is also mostly
included in the Monte Carlo calculation, but since the radiative $b$ decay
leaves four tracks and the reference decay only three, we accounted for
the second meson track by multiplying the Monte Carlo efficiency by
the integrated CTC tracking efficiency, $0.96\pm 0.02$, estimated by
embedding simulated tracks in CDF data~\cite{Abe:1998andreas} (see Section~\ref{mi:eff}).
As previously noted, the Monte Carlo simulation already models the SVX
efficiency, and thus no further correction to the tracking efficiency
is needed.

Effects which do not cancel in the ratio include the efficiencies of
the quality cuts for the $J/\psi$ partner electron, the
$\Lambda\rightarrow p\pi^-$ selection, and the conversion probabilities.

The quality cut efficiency of the $J/\psi$ partner electron was estimated
from the $J/\psi$ candidates themselves to be $0.75\pm 0.03$
by counting the number of the $J/\psi$ signals before and after the 
quality cut.
 In a similar manner, the $\Lambda$ quality cut efficiency was estimated to
be $0.72\pm 0.02$.  We investigated the effect of the photon conversion
probability in detail because it dominates the total efficiency differences
between the radiative $b$ decays and the reference decay.

The detector simulation, described in Section~\ref{mi:eff}, also simulates
photon conversions.  The material distribution of the CDF inner detector
used by the simulation is based on previous photon conversion measurements
and a careful accounting of the material of the CTC inner wall which is
known to be $(1.26 \pm 0.06)\%$ of a radiation length.
We calibrated the simulation by normalizing the conversions simulated 
in the CTC inner wall with the rate seen in the data.
The data used
consists of the $\overline{B}_d^0\rightarrow\overline{K}^{*0}\gamma$
candidates, but with loose selection cuts on $ct$, $I_B$, and mass to
increase the sample size.  
The resulting conversion probability from the
Monte Carlo calculations is $\sim 6\%$.  
The simulation was analyzed in the same manner
as the data; in this way, the non-uniformity in the material distribution
and the consequent dependence of the conversion probability on the physics
process and event selection criteria was included in the simulation 
calibration.
In particular, requiring the meson tracks to be reconstructed in the SVX,
as is the case in the $\overline{B}_d^0$ and $\overline{B}_s^0$ samples,
implies that most of the photons will pass through approximately $1\%X_0$
more material than those in events where the tracks lie outside the SVX 
fiducial
volume.  On the other hand, the $\Lambda_b^0$ analysis makes no SVX
requirements on the tracks; since 50\% of such photons are outside the
SVX volume, they traverse, on average, $\sim 0.5\%X_0$ less material
compared to the $B$ meson case.  The process dependent scale factors 
which relate the data samples to the simulation normalization
are found to be $0.89\pm 0.05$ for the $\overline{B}_d^0$ and 
$\overline{B}_s^0$ decays, and $0.95\pm 0.05$ for the $\Lambda_b^0$ decay.

Table~\ref{Table:EfficienciesConv} shows a summary of the efficiency 
estimates for each of the decay modes.  For example, the ratio for
$\overline{B}_d^0\rightarrow\overline{K}^{*0}\gamma$ is given by
$0.064\times 0.89\times (0.96/0.75)$, where 0.064 is the Monte Carlo
efficiency ratio, 0.89 is the conversion probability scale factor,
0.96 is the CTC tracking efficiency for the second meson track, and
0.75 is the partner electron quality cut efficiency for the
$J/\psi\rightarrow e^+e^-$ decay in the reference sample.  As expected,
the efficiency ratio is around 6\%, largely due to the conversion
probability.  

The single event sensitivities defined by 
Eqs.~\ref{Eq:Sensitivity} 
and \ref{Eq:Kstg}--\ref{Eq:Lamg}  are also shown in
Table~\ref{Table:EfficienciesConv}.  They are $4.4\times 10^{-5}$ for $\overline{B}_d^0$,
$9.5\times 10^{-5}$ for $\overline{B}_s^0$, and $2.8\times 10^{-4}$
for $\Lambda_b^0$.  The differences among the sensitivities are
dominated by the differences among the $b$ quark fragmentation fractions.

\subsection{Systematic uncertainties}

Table~\ref{Table:SystamticsConv} summarizes the sources of systematic 
uncertainties
for each of the decay modes considered in this analysis.  One of the
largest uncertainties arises from the statistical uncertainty in the
$J/\psi K$ yield, contributing 21\% for $\overline{B}_d^0$,
18\% for $\overline{B}_s^0$, and 22\% for the $\Lambda_b^0$ channel.
The uncertainty due to the input branching fractions is dominated by
that of ${\cal B}(B_u^-\rightarrow J/\psi K^-)$, and we assign it 11\%
for all the decay modes.

The other major source of systematic uncertainty is the measurement
of the fragmentation fractions $f_s/f_u$ and
$f_{\Lambda_b^0}/f_u$~\cite{Affolder:2000iq}.
These fractions were measured at CDF using the decays
$\overline{B}_s^0\rightarrow e^-D_s^+X$ and
$\Lambda_b^0\rightarrow e^-\Lambda_c^+X$, normalized to
$B_u^-\rightarrow e^-D^0X$.  Their quoted uncertainties are 18\% for
$f_s/f_u$ and 35\% for $f_{\Lambda_b^0}/f_u$, but these values
include a 6\% uncertainty, originating from the $b$ hadron $p_T$
spectrum, which is fully correlated with the corresponding
uncertainty in this analysis.  
We thus reduced the quoted uncertainties
by 6\% in quadrature and obtained a 17\% systematic uncertainty due to
$f_s/f_u$ and 34\% due to $f_{\Lambda_b^0}/f_u$.

We confirmed that changing the $b$ quark $p_T$ spectrum does not contribute
any systematic uncertainty, since this spectrum is common to all the
decay modes, by changing the Monte Carlo generation parameters from
their nominal values $m_b=4.75\;{\rm GeV}/c^2$ and $\mu = \mu_0$.  The
$b$ quark mass was changed to 4.5 and $5.0\;{\rm GeV}/c^2$, and the
renormalization scale was changed to $\mu_0/2$ and $2\mu_0$.  Individual
efficiencies for the radiative and $B_u^-\rightarrow J/\psi K^-$ decays
vary by $\sim 20\%$, but the efficiency ratios remain, as expected,
stable within the uncertainties of the finite Monte Carlo samples.

Small systematic uncertainties are contributed by efficiency factors which
do not cancel in the ratio.  For instance, for
the photon conversion probability correction, which was
evaluated to be $0.89\pm 0.05$ for the $B$ mesons, we assign a 6\% systematic
uncertainty. For the $\Lambda_b^0$ case, the uncertainty is 5\%.
We assign a 4\% systematic uncertainty for the quality cut efficiency on
the partner electron in the $J/\psi\rightarrow e^+e^-$ decay, and 3\%
for the quality cut efficiency for reconstructing $\Lambda\rightarrow p\pi^-$.
These two uncertainties arise from the data sample sizes used for
the efficiency estimation.  The CTC tracking efficiency
contributes another 2\% systematic uncertainty
which comes from its instantaneous luminosity and electric charge dependence.

Another effect which does not cancel in the efficiency ratio is 
that the hadronic/electromagnetic energy ratio cut depends on the
number of tracks pointing to the calorimeter cluster.  This number is
different for photon conversions and $J/\psi\rightarrow e^+e^-$
decays. About 45\% of the conversion partners point to the same
cluster as the trigger electron, while less than 1\% of the partner electrons
in $J/\psi$ decay exhibit the same behavior.  In principle, the effect
of this difference can be estimated with a full simulation of the
$p\overline{p}$ event, including $b$ fragmentation products and multiple
$p\overline{p}$ collisions.  Instead, we estimated this systematic uncertainty
to be about 5\% based on the efficiency difference between the two different
hadronic/electromagnetic energy ratio cuts on the $J/\psi\rightarrow e^+e^-$
candidates in the data.

Finally, the systematic uncertainties due to the finite Monte Carlo sample
sizes in the efficiency calculations were all around 4\%.  When all these
uncertainties were combined in quadrature, we found the total systematic
uncertainties to be 26\% for $\overline{B}_d^0$, 29\% for
$\overline{B}_s^0$, and 43\% for $\Lambda_b^0$.

\subsection{Results}

The low background level for $\overline{B}_d^0$ and $\overline{B}_s^0$
radiative decays allows us to set limits on the branching fractions without
background subtraction.  For the $\Lambda_b^0$ case, however, we account for
the expected background level by using a simple simulation which generates
the numbers of signal and background events in each trial according to
the probability distributions ${\cal P}_{N_{{\rm CL}},\sigma_{{\rm syst}}}(n)$
and ${\cal P}_{N_{{\rm bg}},\sigma_{{\rm bg}}}(n)$, where
${\cal P}_{\mu,\sigma}(n)$ is defined in Eq.~\ref{Eq:PG}. $N_{{\rm CL}}$ is
the upper limit on the number of decays for a given CL, $\sigma_{{\rm syst}}$
is the systematic uncertainty on the signal yield, and $N_{{\rm bg}}$ is
the number of background events with uncertainty $\sigma_{{\rm bg}}$.  The
CL is given by the fraction of trials which has the total number of signal
and background events exceeding the observed number of events $N_{{\rm obs}}$,
but still has fewer background events than $N_{{\rm obs}}$.

We calculated $N_{{\rm CL}}$ to be 4.3 for $\overline{B}_d^0$, 2.6 for
$\overline{B}_s^0$, and 4.5 for $\Lambda_b^0$ at 90\% CL, and
5.5, 3.5, and 6.8, respectively, at 95\% CL.  
With the single event sensitivities
listed in Table~\ref{Table:EfficienciesConv}, 
we obtained the limits on the branching fraction,
$\Br(\Btogkst) <$ $1.9 \times 10^{-4}$ ($2.4 \times 10^{-4}$),
$\Br(\Btogphi) <$ $2.5 \times 10^{-4}$ ($3.4 \times 10^{-4}$),
and $\Br(\Lambtoglam) <$ $ 1.3 \times 10^{-3}$ ($1.9 \times 10^{-3}$)
at 90\% (95\%) CL.
%

%
\section{Combined Limits}

Since the two analyses searching for
$\overline{B}_d^0\rightarrow\overline{K}^{*0}\gamma$ and
$\overline{B}_s^0\rightarrow\phi\gamma$ decays are statistically
independent, we simply add the numbers of candidates found in each
analysis.  In total, there are two $\overline{B}_d^0$ candidates with
an expected background of $0.6\pm 0.3$ events, and no $\overline{B}_s^0$
candidates with an expected background of $0.1\pm 0.1$ events.  The
combination does not yield any significant excesses over the background
level but does tighten the upper limits on the branching fractions.

The combined single event sensitivity of using both methods is given by
$S_{I+II}^{-1} = S_I^{-1} + S_{II}^{-1}$ and is $2.0\times 10^{-5}$
for $\overline{B}_d^0$ and $4.7\times 10^{-5}$ for $\overline{B}_s^0$.
The systematic uncertainties due to the generated $p_T(B)$ spectrum,
$f_s/f_u$, ${\cal B}(\phi\rightarrow K^+K^-)$, and CTC pattern recognition
efficiency are fully correlated between the two methods and simply added
together; the other systematic uncertainties are considered to be fully
uncorrelated and are thus added in quadrature.  We obtained 18\% as the
combined systematic uncertainty for $\overline{B}_d^0$ and 25\% for
$\overline{B}_s^0$.  We then calculated, without any background
subtraction, the upper limits on the branching fractions
$\Br(\Btogkst) <$ $1.1 \times 10^{-4}$ ($1.4 \times 10^{-4}$) and
$\Br(\Btogphi) <$ $1.2 \times 10^{-4}$ ($1.6 \times 10^{-4}$)
at 90\% (95\%) CL.
%

%
\section{Conclusions}

We have searched for
$\bd \to \kst(\to K^-\pi^+)\gamma$, 
$\bs \to \phi(\to K^+K^-)\gamma$, 
$\lamb \to \Lambda(\to p\pi^-)\gamma$, 
and their charge conjugate decays,
using events produced in \ppbar\ collisions at $\sqrt{s}=1.8$ TeV 
and recorded by CDF.
Two methods were employed. 

In the first method, 
the photon was detected in the electromagnetic 
calorimeter as a cluster of energy.
We designed and installed a dedicated trigger which,
in addition to the photons, 
required information about the charged particles 
originating from the daughter meson.
We collected 22.3 \invpb\ of data 
with $E_T(\gamma) > 10$ GeV during 1995
and 6.6 \invpb\ of data 
with $E_T(\gamma) > 6$ GeV during 1995--96.

In the second method, 
the photon was identified by an electron-positron pair
produced through external photon conversion before the
tracking detector volume. 
One of the conversion electrons with $E_T > 8$ GeV
served as a trigger for event recording;
no additional tracks coming from the daughter hadron decay were required.
The trigger recorded 74 pb$^{-1}$ of data from the 1994--96 period.
We observed no significant signal in both the methods, and set upper 
limits on the branching fractions (Table~\ref{Table:Results}).

Combining the two analyses, we obtained upper limits on the branching fractions
    \begin{eqnarray}
\nonumber
    {\cal B}(\overline{B}_d^0\rightarrow\overline{K}^{*0}\gamma) & < &
            1.4\times 10^{-4} \\
\nonumber
    {\cal B}(\overline{B}_s^0\rightarrow\phi\gamma) & < & 1.6\times 10^{-4} \\
\nonumber
    {\cal B}(\Lambda_b^0\rightarrow\Lambda\gamma) & < & 1.9\times 10^{-3}
    \end{eqnarray}
    at 95\% CL. 
The result on the $\overline{B}_d^0\rightarrow\overline{K}^{*0}\gamma$ decays
is consistent with the measurements performed in the $e^+ e^-$ colliders
\cite{Coan:2000kh,Convery:2001pc,Ushiroda:2001sb}.
The results on the $\overline{B}_s^0$ and $\Lambda_b$ decays
are the current lowest limit on these branching fractions and
they are also consistent with the theoretical prediction that the 
$\overline{B}_s^0\rightarrow\phi\gamma$ and 
$\overline{B}_d^0\rightarrow\overline{K}^{*0}\gamma$ 
branching fractions are of the same magnitude~\cite{Ali:1994excl}.

\section*{Acknowledgments}

We thank the Fermilab staff and the technical staffs of the
participating institutions for their vital contributions.
This work was supported by the U.S. Department of Energy
and the National Science Foundation; 
the Natural Sciences and Engineering Research Council of Canada; 
the Istituto Nazionale di Fisica Nucleare of Italy; 
the Ministry of Education, Science, Sports and Culture of Japan; 
the National Science Council of the Republic of China; 
and the A.~P.~Sloan Foundation.
%


%

%

%
\onecolumn
%

%
%


\begin{figure}
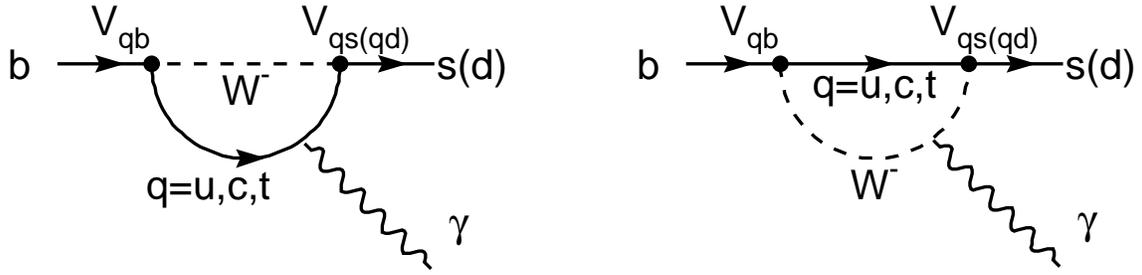

\caption{Feynman diagrams for the $b \to s \gamma$ and $b \to d \gamma$
penguin loops.}
\end{figure}

\begin{figure}
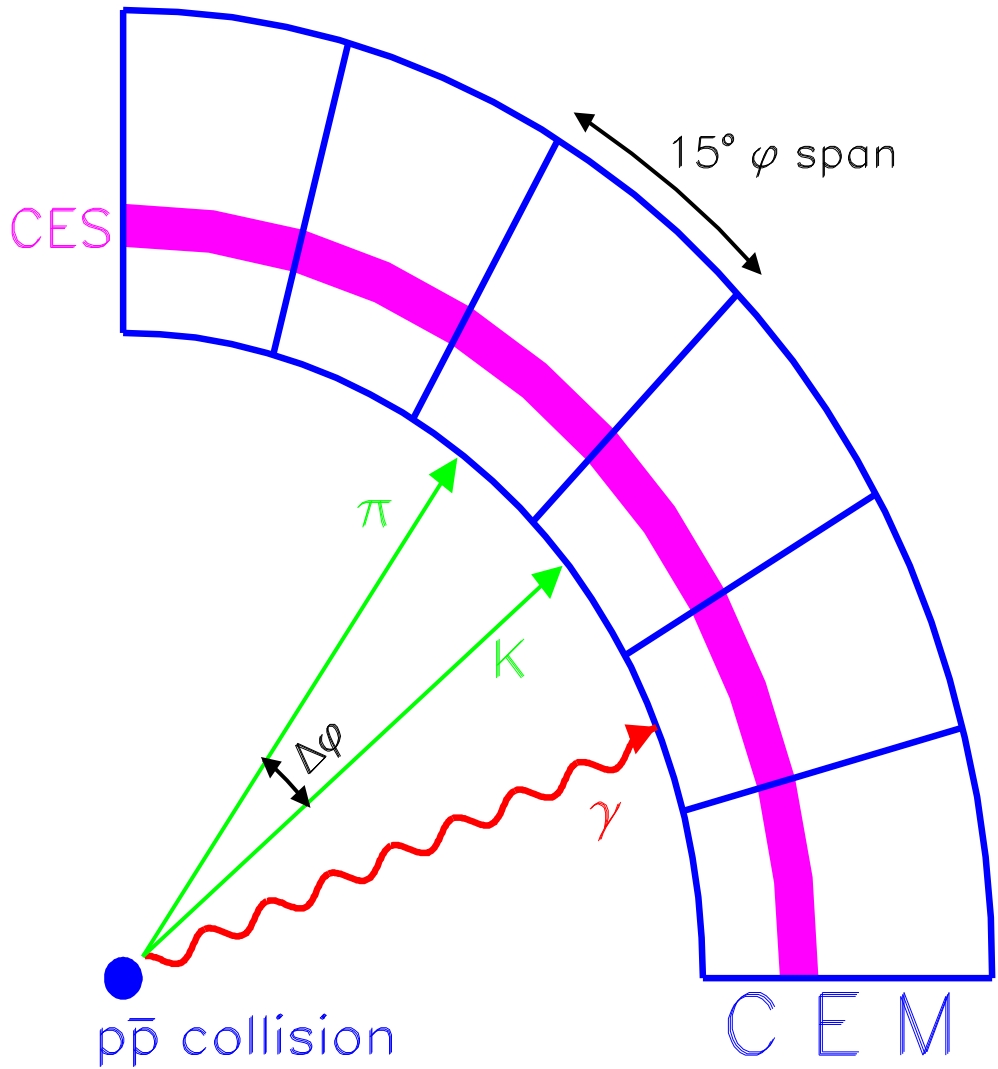

\caption{The topology of the objects considered by the penguin trigger, shown 
on a schematic depiction of the CEM calorimeter with the beam pipe going 
perpedicularly through this page.}
\end{figure}

\begin{figure}
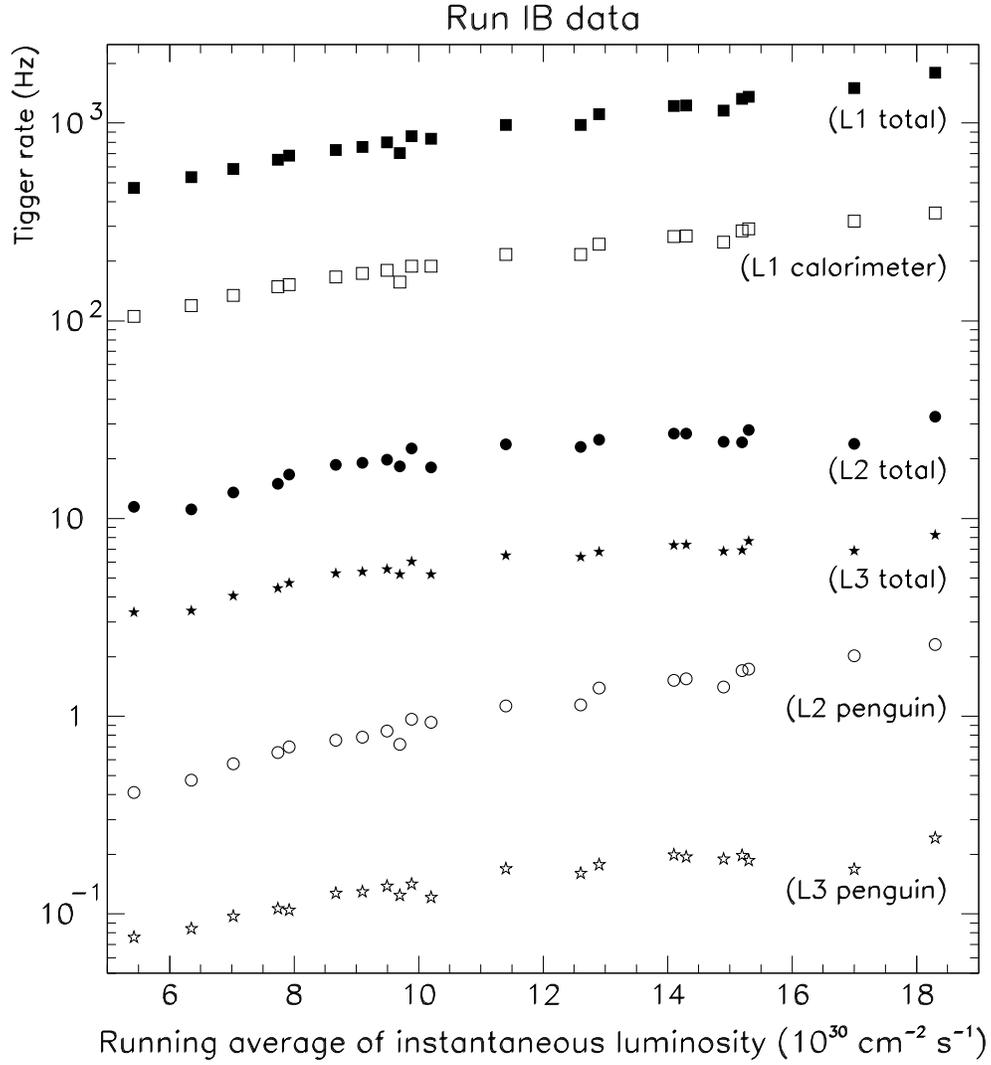

\caption{Level 1, 2, and 3 trigger rates for the photon + 2 track trigger
as a function of instantaneous luminosity in Run IB (open points).
Total trigger rates for each stage are also shown (filled points).}
\end{figure}

\begin{figure}
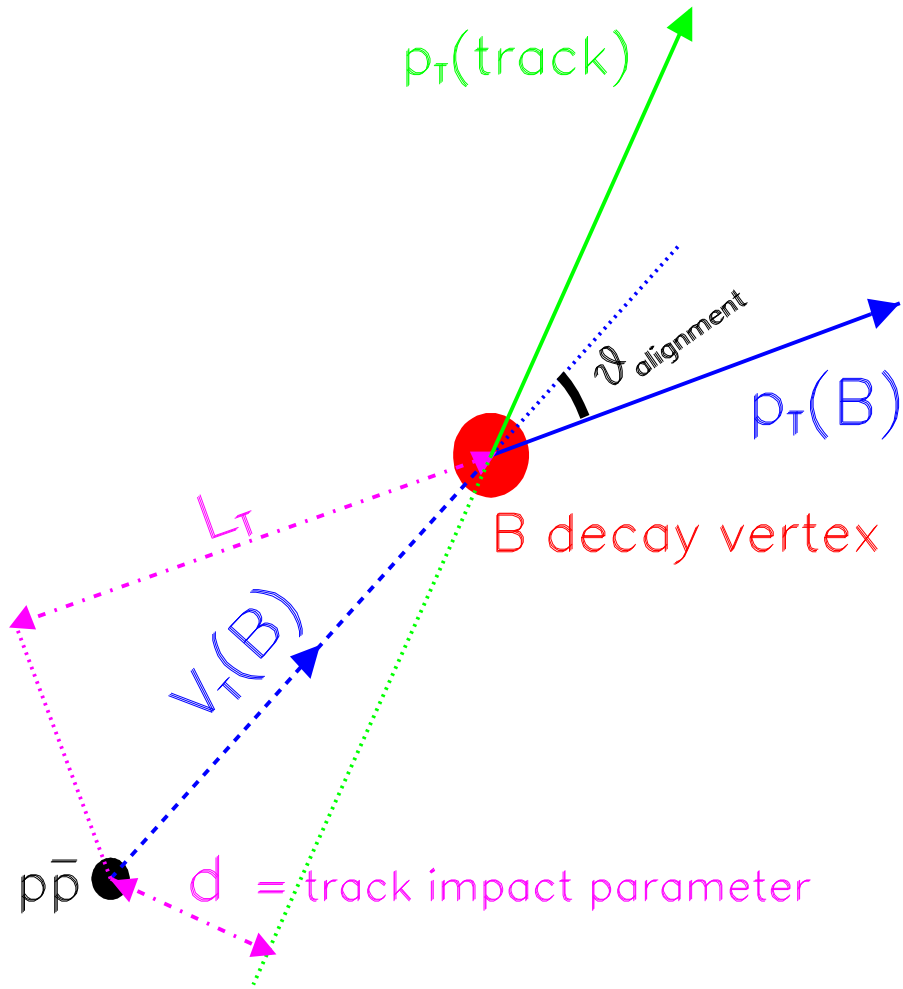

\caption{The $B$ decay vertex and relevant quantities on the plane transverse to the beam. For clarity, only the $B$ momentum and one of its' charged daughters are shown.}
\end{figure}


\begin{figure}
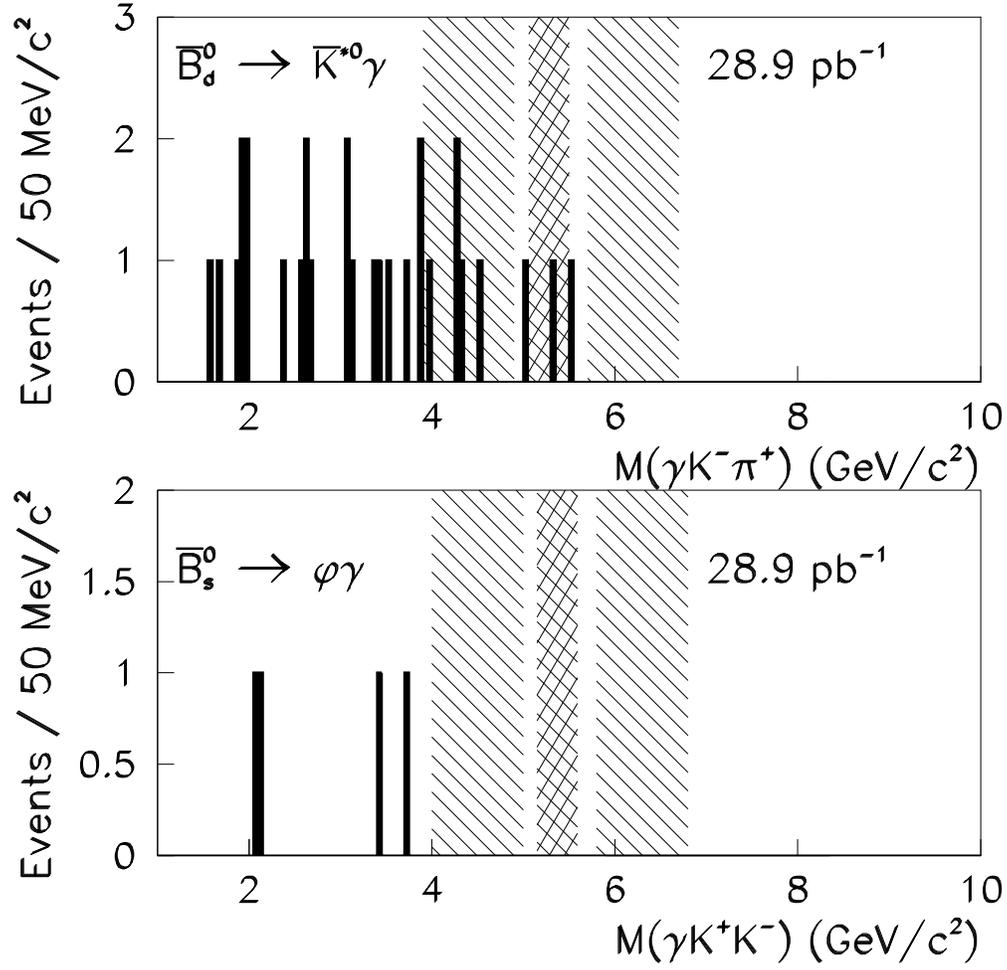

\caption{ Top:  $\gamma K^-\pi^+$ invariant mass distribution
for 
$\overline{B}_d^0 \rightarrow \gamma \overline{K}^{*0}(\rightarrow K^- \pi^+)$.
There is one candidate.
Bottom:  $\gamma K^+K^-$ invariant
mass distribution for 
$\overline{B}_s^0 \rightarrow \gamma \phi(\rightarrow K^+ K^-)$.  
There are no candidates seen.
}
\end{figure}

\begin{figure}
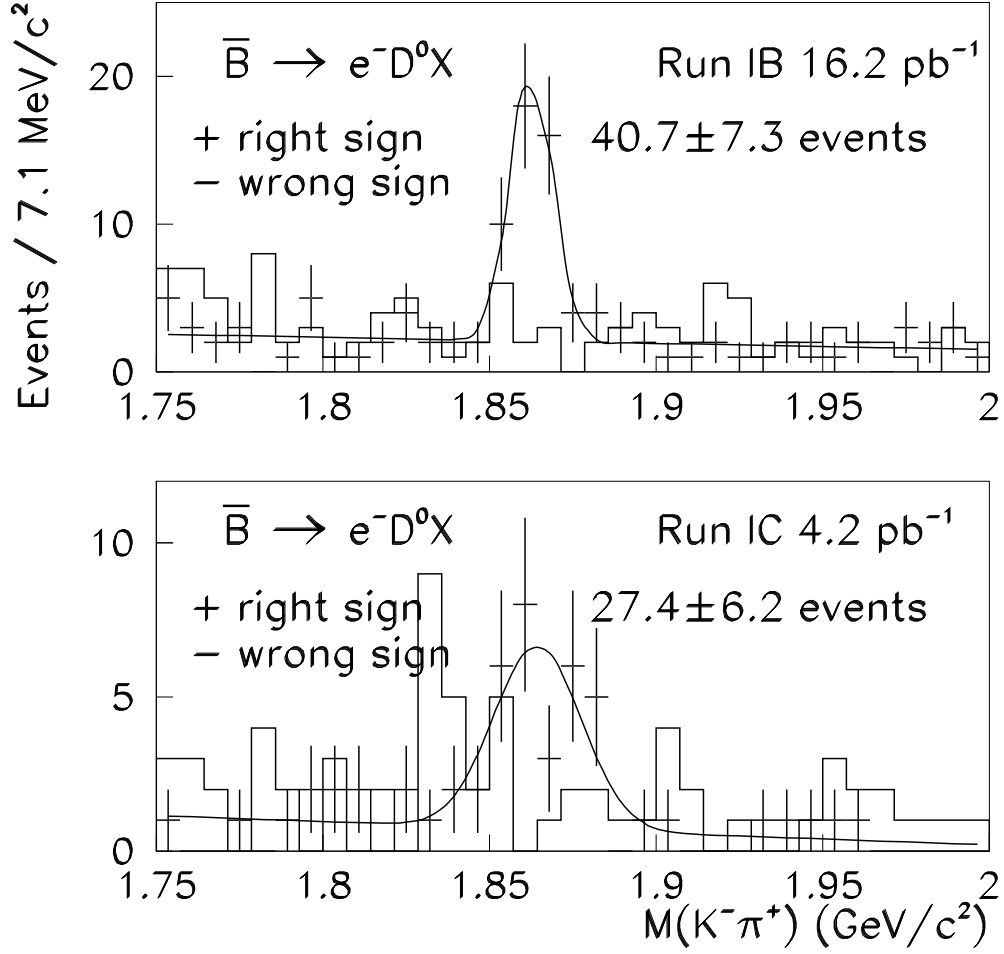

\caption{
Invariant mass distributions of the $K^{-}\pi^{+}$ combinations 
for 
$\overline{B}\rightarrow e^- D^0(\rightarrow K^- \pi^+) X$,
decays in the Run \ib\ (top) and \ic\ (bottom) data.
   The right-sign distributions (points) are for same
   charge electrons and kaons, as should be the case if they are both
   products of the real $B$ decay chain, whereas in the wrong-sign
   distributions (histograms) the kaon has opposite charge to the electron.
   By fitting a Gaussian and a straight line to the right-sign distributions 
   we find
   $40.7 \pm 7.3$ and $27.4 \pm 6.2$
   candidate $\overline{B}\rightarrow e^- D^0(\rightarrow K^- \pi^+) X$
   events in Runs \ib\ and \ic, respectively.}
\end{figure}


 \begin{figure}  
\caption{Photon conversion vertex density  in the $x-y$ plane 
in the 74 pb$^{-1}$ of  CDF Run IB inclusive electron data.
The fine structure of the CDF tracking detectors
can be clearly resolved.}
\end{figure}

 \begin{figure}  
\caption{Photon conversion vertex density  in the $r-z$ plane 
in the 74 pb$^{-1}$ of  CDF Run IB inclusive electron data.
The fine structure of the CDF tracking detectors
can be clearly resolved.}
\end{figure}

 \begin{figure}
\caption{
Top:  $e^+e^-K^-\pi^+$ invariant mass distribution for
$\overline{B}_d^0 \rightarrow \overline{K}^{*0}(\rightarrow K^- \pi^+) \gamma(\rightarrow e^+ e^-)$
in the $74\;{\rm pb}^{-1}$ of CDF Run IB inclusive electron data.
Bottom:  corresponding $e^+e^-K^-$ invariant mass
distribution for the $B_u^-\rightarrow J/\psi(\rightarrow e^+ e^-) K^-$
reference decay.  There are $28.0\pm 5.8$ events after
background subtraction.
}
\end{figure}

 \begin{figure}
\caption{
Top:  $e^+e^-K^+K^-$ invariant mass distribution for
 $\overline{B}_s^0 \rightarrow \phi(\rightarrow K^+ K^-) \gamma(\rightarrow e^+ e^-)$ in
                 the $74\;{\rm pb}^{-1}$ of CDF Run IB inclusive electron
                 data.
                 Bottom:  corresponding $e^+e^-K^-$ invariant mass
                 distribution for the 
                 $B_u^-\rightarrow J/\psi(\rightarrow e^+ e^-) K^-$
                 reference decay.  There are $35.0\pm 6.4$ events after
                 background subtraction.
}
 \end{figure}

 \begin{figure}
\caption{
Top:  $e^+e^-p\pi^-$ invariant mass distribution for
  $\Lambda_b^0 \rightarrow \Lambda(\rightarrow p \pi^-) \gamma(\rightarrow e^+ e^-)$ in
                 the $74\;{\rm pb}^{-1}$ of CDF Run IB inclusive electron
                 data.  
                 Bottom:  corresponding $e^+e^-K^-$ invariant mass
                 distribution for the 
                 $B_u^-\rightarrow J/\psi(\rightarrow e^+ e^-) K^-$
                 reference decay.  There are $24.0\pm 5.3$ events after
                 background subtraction.
}
 \end{figure}

 \begin{figure}
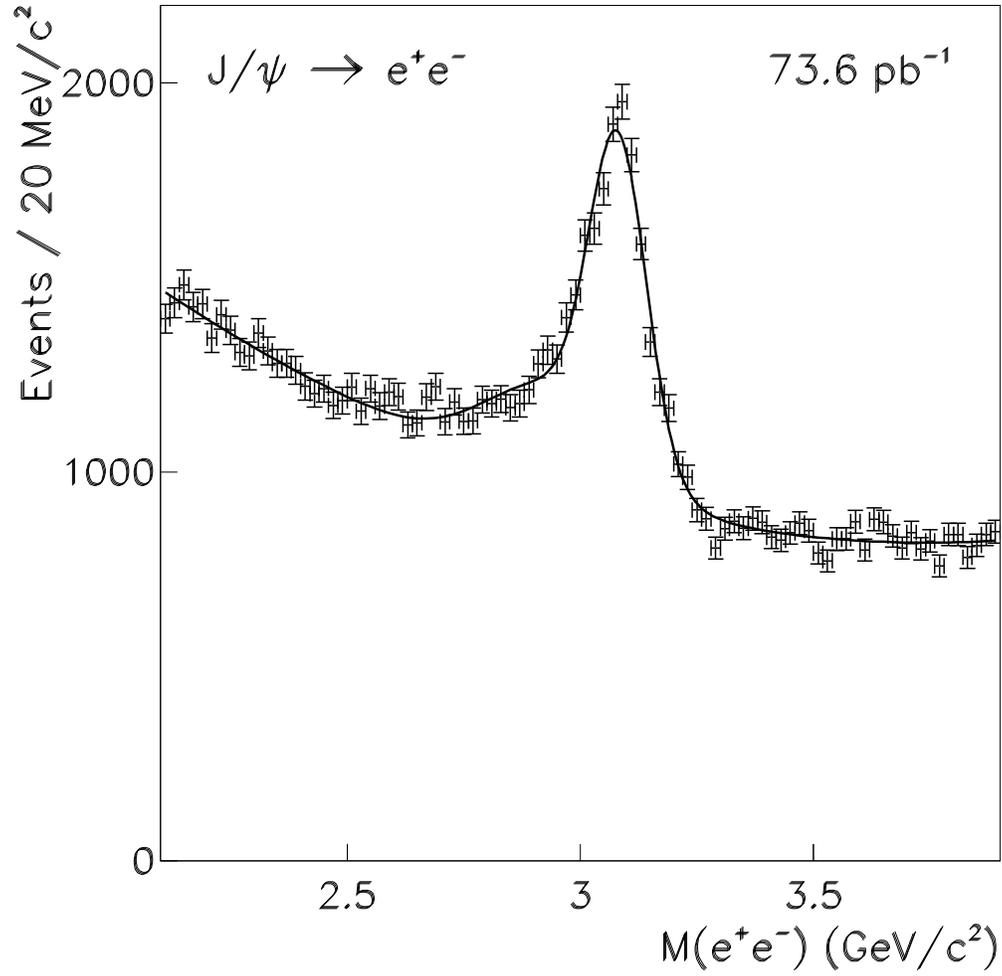
  
\caption{Dielectron invariant mass distribution of the $J/\psi \to e^+e^-$
candidates in the 74 pb$^{-1}$ of  CDF Run IB inclusive electron data. 
The number of the $J/\psi \to e^+e^-$ events obtained by fitting
the mass distribution to a function of 2 Gaussians and a polynomial
is $\sim$ 8000.}
\label{Fig:Jpsiee}
\end{figure}

%
%
%

\newpage

\begin{table}
\caption{ 
  Ingredients for the calculation of the branching fractions of
  \Btogkst\ and \Btogphi\ decays in Method I
  according to Eqs.~(\ref{Eq:Bd_peng_Brel}) and (\ref{Eq:Bs_peng_Brel}).
}
\label{Table:BRrel_BC_Peng}
\begin{center}
\begin{tabular}{lcccc} 
   & 
   \multicolumn{2}{c}{\Btogkst} &
   \multicolumn{2}{c}{\Btogphi} \\
\cline{2-5}
    & Run \ib & Run \ic & Run \ib & Run \ic \\ 
\hline
$N_{\rm obs}$ (events)& 
        $0$ & $1$ & $0$ & $0$ \\
$N_{\rm bg}$ (events)& 
        $0.9 \pm 0.4$ & $0.2 \pm 0.2$ 
        & $<0.54$ (90\% CL) & $< 0.54$ (90\% CL) \\
$N_{\edz}$ (events)& 
        $40.7 \pm  7.3$ & $27.4 \pm 6.2$ & $40.7 \pm  7.3$ & $27.4 \pm 6.2$ \\
\\
$\left. f_{d,s} \right/(f_u+f_d)$
         & \multicolumn{2}{c}{1/2} & \multicolumn{2}{c}{$0.213 \pm 0.038$} \\ 
$\Br(\Ksttokpi)$  
        &  \multicolumn{2}{c}{$2/3$}  &  \multicolumn{2}{c}{---} \\
$\Br(\Phitokk)$
       &  \multicolumn{2}{c}{---} &  \multicolumn{2}{c}{$0.491 \pm 0.008$} \\ 
$\Br(\Btoedx) \cdot \Br(\Dtokpi)$ 
       & \multicolumn{2}{c}{$(2.94 \pm 0.4) \times 10^{-3}$ }
       & \multicolumn{2}{c}{$(2.94 \pm 0.4) \times 10^{-3}$ } \\
\\
$\left. \epsilon_{\rm signal} \right/ \epsilon_{\edz}$ & 
            2.65  &       2.01 &       3.50  &       2.48 \\ 
$\left. L_{\rm peng}\right/L_{eX}$ &
        $22.3/16.2$ & $6.6/4.2$ & $22.3/16.2$ & $6.6/4.2$ \\ 
\hline
Single event sensitivity   & $5.94 \times 10^{-5}$  & $10.2  \times 10^{-5}$
              & $1.44  \times 10^{-4}$ & $2.64  \times 10^{-4}$ \\
Combined     & \multicolumn{2}{c}{$3.75 \times 10^{-5}$}
              & \multicolumn{2}{c}{$9.29 \times 10^{-5}$} \\
\end{tabular}
\end{center}
\end{table}

\begin{table}
\caption{ 
  Systematic uncertainties on the branching fractions of 
  \Btogkst\ and \Btogphi\ decays in Method I.
}
\label{Table:syst_unc_BC_Peng}
\begin{center}
\begin{tabular}{lcccc} 
&\multicolumn{2}{c}{\Btogkst} & \multicolumn{2}{c}{\Btogphi} \\ 
Source & Run \ib & Run \ic & Run \ib & Run \ic \\ 
\hline
\Edz\ statistics  
            & $18\%$ & $23\%$ & $18\%$ & $23\%$ \\
Monte Carlo statistics  
              & $2\%$ & $2\%$ & $2\%$ & $2\%$ \\     
Composition of $e+D^0$ sample                     
            & $12\%$ & $11\%$ & $12\%$ & $11\%$ \\ 
$p_T(B)$ distribution                         
            & $3\%$ & $3\%$ & $5\%$ & $2\%$ \\ 
CEM $E_T$ cut efficiency       
            & $7\%$ & $7\%$ & $8\%$ & $8\%$ \\                                
CFT efficiency    &  $3\%$ & $3\%$ & $3\%$ & $3\%$ \\
CTC pattern recognition & $2\%$   & $2\%$   & $2\%$   & $2\%$ \\   
XCES efficiency   &  $2\%$  & $2\%$  & $2\%$  & $2\%$ \\
Isolation efficiency  & $2\%$  & $2\%$  & $2\%$  & $2\%$ \\
$\left.f_{s}\right/(f_u+f_d)$  
                  & \multicolumn{2}{c}{---} &\multicolumn{2}{c}{$18\%$} \\ 
$\Br(\Btoedx) \cdot \Br(\Dtokpi)$ 
                  & \multicolumn{2}{c}{$14\%$} & \multicolumn{2}{c}{$14\%$} \\ 
$\Br(\Phitokk)$ &  \multicolumn{2}{c}{---} & \multicolumn{2}{c}{$2\%$} \\ 
\hline
Total systematic uncertainty
			& $27\%$ & $30\%$  & $33\%$ & $36\%$ \\
Combined  
			& \multicolumn{2}{c}{25\%} &\multicolumn{2}{c}{31\%} \\
\end{tabular}
\end{center}
\end{table}

\newpage

%
%
 \begin{table}
 \caption{Ingredients for the calculation of the 
branching fractions of \Btogkst, \Btogphi, and \Lambtoglam\ decays in
Method II according to Eqs.~(\ref{Eq:Kstg})--(\ref{Eq:Lamg}).
}
 \label{Table:EfficienciesConv}
\begin{center}
\begin{tabular}{lccc}

 				&\Btogkst
 				&\Btogphi
 				&\Lambtoglam \\
\hline
$N_{\rm obs}$ (events)		&1		&0		&2\\
$N_{\rm bg}$ (events)		&$0.6 \pm 0.3$	&$0.1 \pm 0.1$	&$3.4 \pm 0.6$\\
$N_{\jpsi K}$ (events)		&$28.0\pm 5.8$ 	&$35.0 \pm 6.4$	&$24.0 \pm 5.3$\\
\\
$\left. f_{d,s,\Lambda_b}\right/f_u$
			&1 	&$0.426 \pm 0.076$	&$0.236 \pm 0.082$\\
$\Br(\Ksttokpi)$	&2/3    & --- & --- \\
$\Br(\Phitokk) $	&---	&$0.491 \pm 0.008$ & --- \\
$\Br(\Lamtoppi)$	&---	&---			&$0.639 \pm 0.005$\\ 
$\Br(\Btojpsik)$	&$(0.99 \pm 0.10) \times 10^{-3}$
                        &$(0.99 \pm 0.10) \times 10^{-3}$
                        &$(0.99 \pm 0.10) \times 10^{-3}$ \\
$\Br(\Jpsitoee)$
                        &$(6.02 \pm 0.19) \times 10^{-2}$
                        &$(6.02 \pm 0.19) \times 10^{-2}$
                        &$(6.02 \pm 0.19) \times 10^{-2}$ \\
\\
CTC tracking            &$0.960 \pm 0.020$
                        &$0.960 \pm 0.020$
                        &$0.960 \pm 0.020$   \\
\Jpsi\ partner electron	&$0.749 \pm 0.028$
                        &$0.749 \pm 0.028$
                        &$0.749 \pm 0.028$ \\
\Lam\ quality cut		&---&---&$0.721 \pm 0.018$ \\
$\left. X_T({\rm DATA})\right/X_T({\rm MC})$	
& $0.889 \pm 0.052$ &  $0.889 \pm 0.052$	& $0.954 \pm 0.047$ \\
\\
$\left[\left.\epsilon_{\rm signal}
\right/\epsilon_{\jpsi K}\right]_{\rm MC}$	&0.0644	&0.0748	&0.0666\\
$\left.\epsilon_{\rm signal}
\right/\epsilon_{\jpsi K}$	&0.0733	&0.0853	&0.0588\\
\hline
Single event sensitivity    &$4.36 \times 10^{-5}$ &$9.54 \times 10^{-5}$
							&$2.80 \times 10^{-4}$\\
\end{tabular}
\end{center}
\end{table}

 \begin{table}
 \caption{Summary of the systematic uncertainties for Method II.}
 \label{Table:SystamticsConv}
\begin{center}
\begin{tabular}{lccc}
 				&\Btogkst
 				&\Btogphi
 				&\Lambtoglam\\
\hline
$\jpsi K$ statistics		&21\%	&18\%	&22\%\\
MC statistics			&4\%	&3\%	&4\%\\
Conversion probability		&6\%	&6\%	&5\%\\
\Jpsi\ partner electron		&4\%	&4\%	&4\%\\
\Lam\  $dE/dx$			&---	&---	&3\%\\
CTC pattern recognition		&2\%	&2\%	&2\%\\
\hadem				&5\%	&5\%	&5\%\\
Fragmentation fractions		&0\%	&17\%	&34\%\\
Branching fractions		&11\% 	&11\%	&11\%\\	
\hline
Total				&26\%	&29\%	&43\%\\
\end{tabular}
\end{center}
\end{table}


 \begin{table}
 \caption{Summary of the branching fraction limits.}
 \label{Table:Results}
\begin{center}
\begin{tabular}{lcccccc}
 				&\multicolumn{2}{c}{\Btogkst}
 				&\multicolumn{2}{c}{\Btogphi}
 				&\multicolumn{2}{c}{\Lambtoglam}\\
Confidence level		&90\%&95\%&90\%&95\%&90\%&95\%\\
\hline
Method I 			&$1.6\times10^{-4}$	&$2.1\times10^{-4}$
				&$2.5\times10^{-4}$	&$3.3\times10^{-4}$
				&--&--\\ 
Method II			&$1.9\times10^{-4}$	&$2.4\times10^{-4}$
				&$2.5\times10^{-4}$	&$3.4\times10^{-4}$
				&$1.3\times10^{-3}$	&$1.9\times10^{-3}$\\
\hline
Combined			&$1.1\times10^{-4}$	&$1.4\times10^{-4}$
				&$1.2\times10^{-4}$	&$1.6\times10^{-4}$
				&$1.3\times10^{-3}$	&$1.9\times10^{-3}$\\
\end{tabular}
\end{center}
\end{table}


\clearpage


\setcounter{figure}{0}
\begin{figure}
\begin{center}
\leavevmode
\epsfxsize=0.95\textwidth
\epsffile
{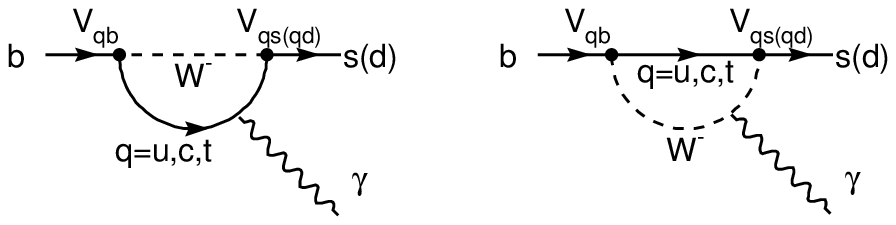}
\end{center}
\caption{Feynman diagrams for the $b \to s \gamma$ and $b \to d \gamma$
penguin loops.}
\label{Fig:PengDiagram}
\end{figure}

\newpage

\begin{figure}
\begin{center}
\leavevmode
\epsfxsize=0.8\textwidth
\epsffile
{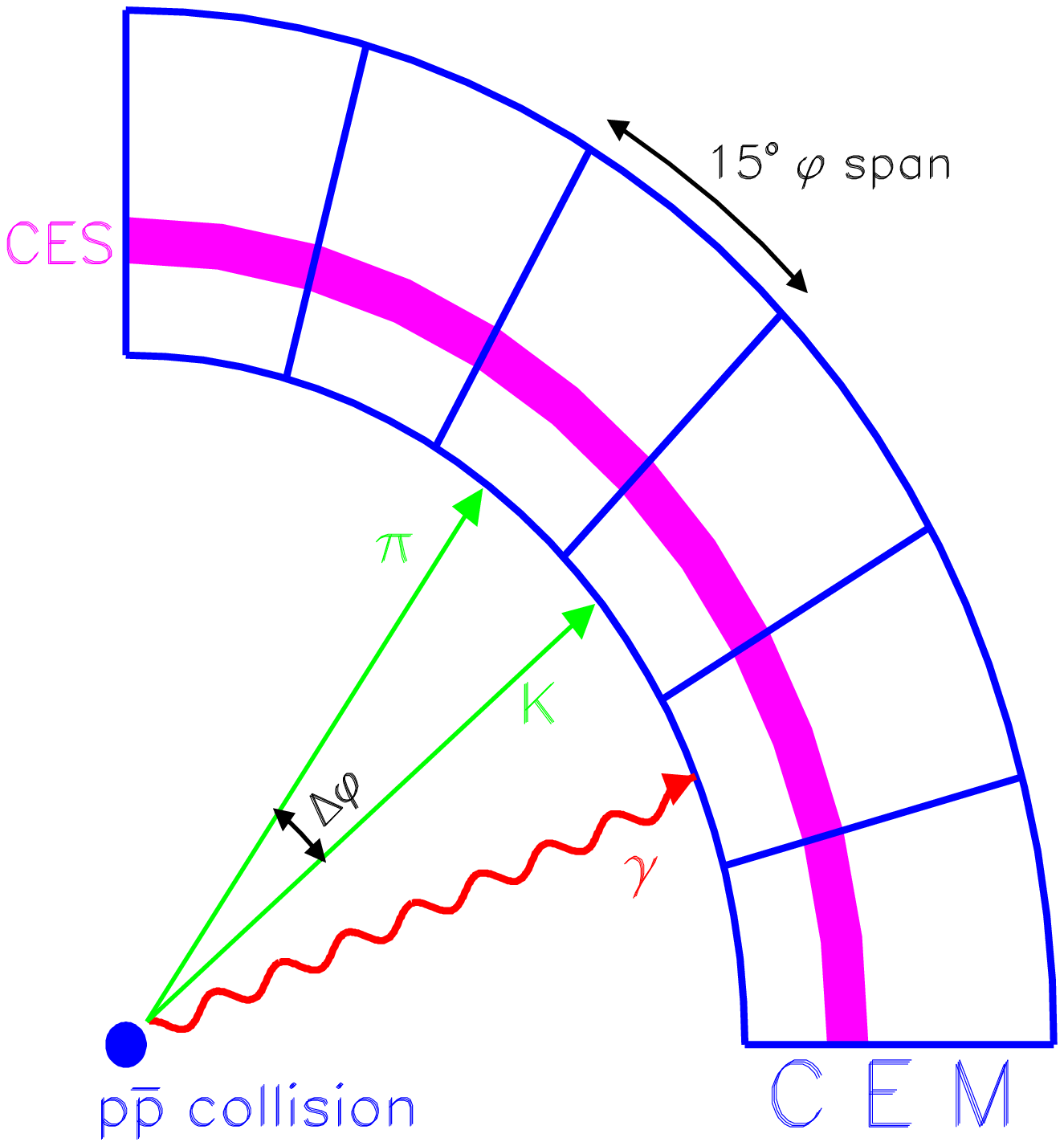}
\end{center}
\caption{The topology of the objects considered by the penguin trigger, shown 
on a schematic depiction of the CEM calorimeter with the beam pipe going 
perpedicularly through this page.}
\label{Fig:PengTrigTopology}
\end{figure}

\begin{figure}
\begin{center}
\leavevmode
\epsfxsize=0.8\textwidth
\epsffile
{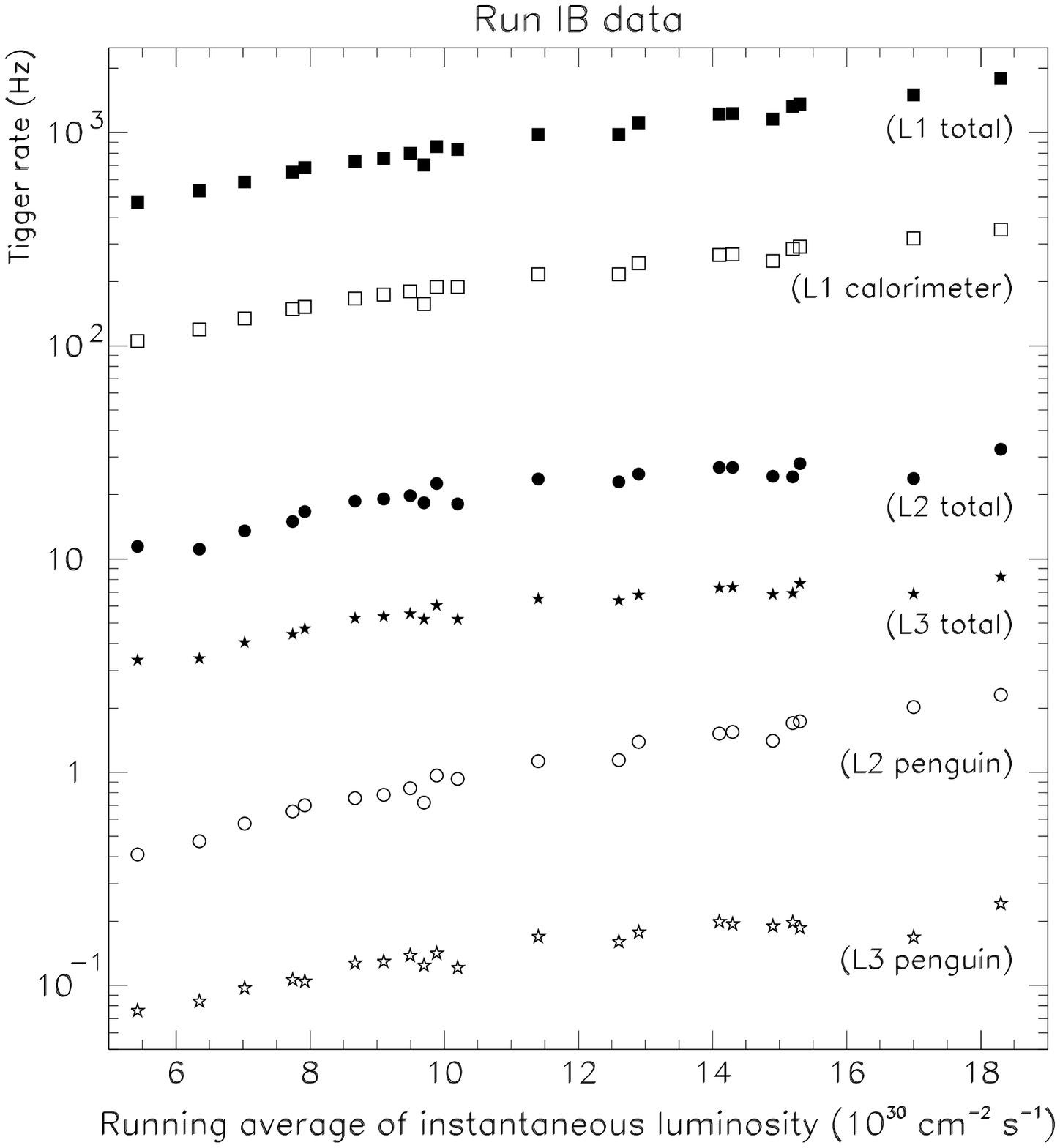}
\end{center}
\caption{Level 1, 2, and 3 trigger rates for the photon + 2 track trigger
as a function of instantaneous luminosity in Run IB (open points).
Total trigger rates for each stage are also shown (filled points).}
\label{Fig:PengTrigRate}
\end{figure}

\begin{figure}
\begin{center}
\leavevmode
\epsfxsize=0.8\textwidth
\epsffile
{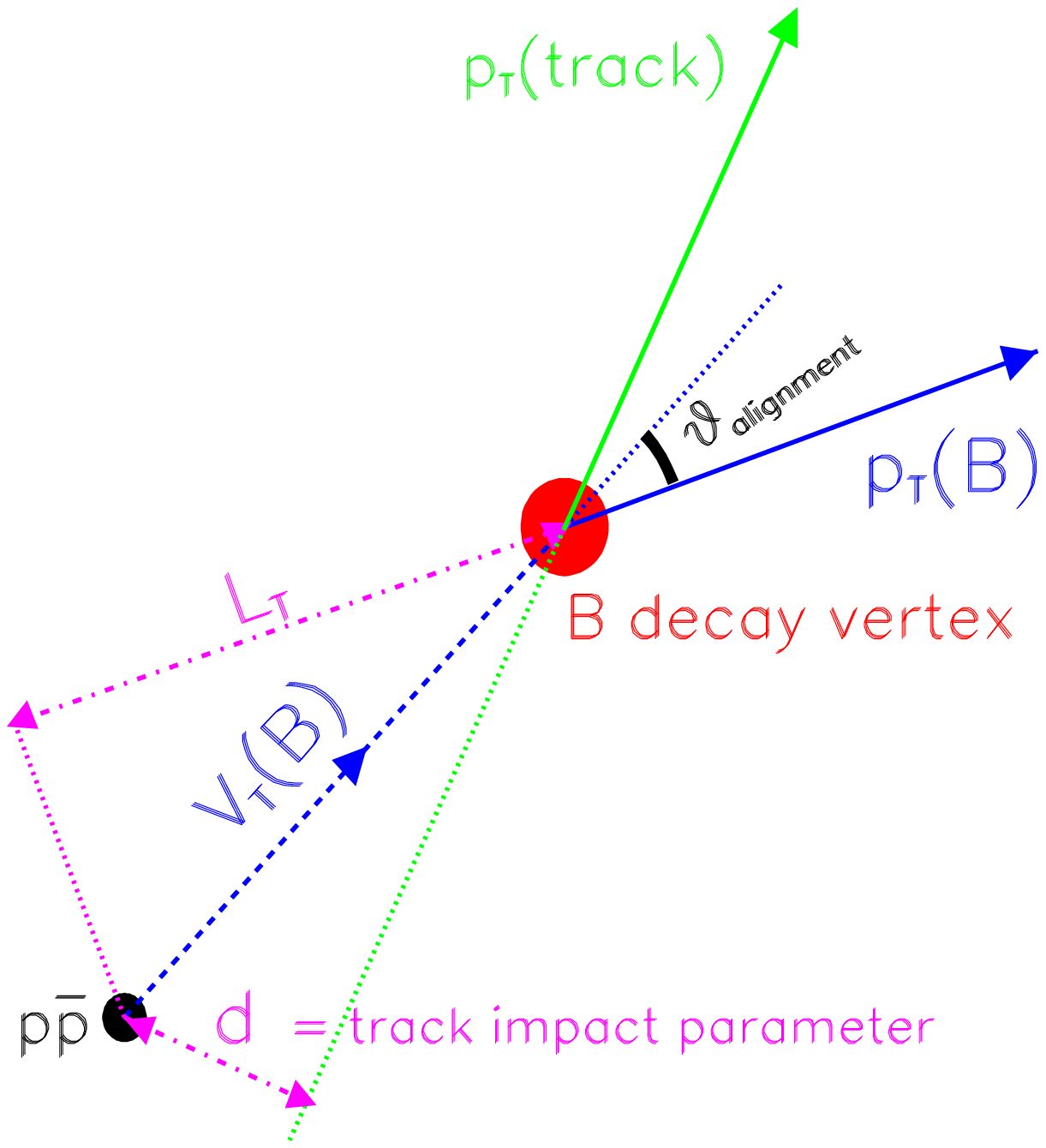}
\end{center}
\caption{The $B$ decay vertex and relevant quantities on the plane transverse to the beam. For clarity, only the $B$ momentum and one of its' charged daughters are shown.}
\label{Fig:Bdecay}
\end{figure}

\begin{figure}
\begin{center}
\leavevmode
\epsfxsize=0.8\textwidth
\epsffile
{radbcand_method1_prd.epsi}
\end{center}
\caption{ 
Top:  $\gamma K^-\pi^+$ invariant mass distribution for 
$\overline{B}_d^0 \rightarrow \gamma \overline{K}^{*0}(\rightarrow K^- \pi^+)$.
There is one candidate.
Bottom:  $\gamma K^+K^-$ invariant
mass distribution for 
$\overline{B}_s^0 \rightarrow \gamma \phi(\rightarrow K^+ K^-)$.
There are no candidates seen.
}
\label{Fig:Bd_Bs_allcuts_Peng}
\end{figure}


\begin{figure}
\begin{center}
\leavevmode
\epsfxsize=0.8\textwidth
\epsffile
{ele_d0.epsi}
\end{center}
\caption{ Invariant mass distributions of the $K^{-}\pi^{+}$ combinations 
for 
$\overline{B}\rightarrow e^- D^0(\rightarrow K^- \pi^+) X$,
decays in the Run \ib\ (top) and \ic\ (bottom) data.
   The right-sign distributions (points) are for same
   charge electrons and kaons, as should be the case if they are both
   products of the real $B$ decay chain, whereas in the wrong-sign
   distributions (histograms) the kaon has opposite charge to the electron.
   By fitting a Gaussian and a straight line to the right-sign distributions 
   we find
   $40.7 \pm 7.3$ and $27.4 \pm 6.2$
   candidate $\overline{B}\rightarrow e^- D^0(\rightarrow K^- \pi^+) X$
   events in Runs \ib\ and \ic, respectively.}
\label{Fig:d0_all_cuts}
\end{figure}

\clearpage

\begin{figure}
\begin{center}
\leavevmode
\epsfxsize=1.0\textwidth
\epsffile
{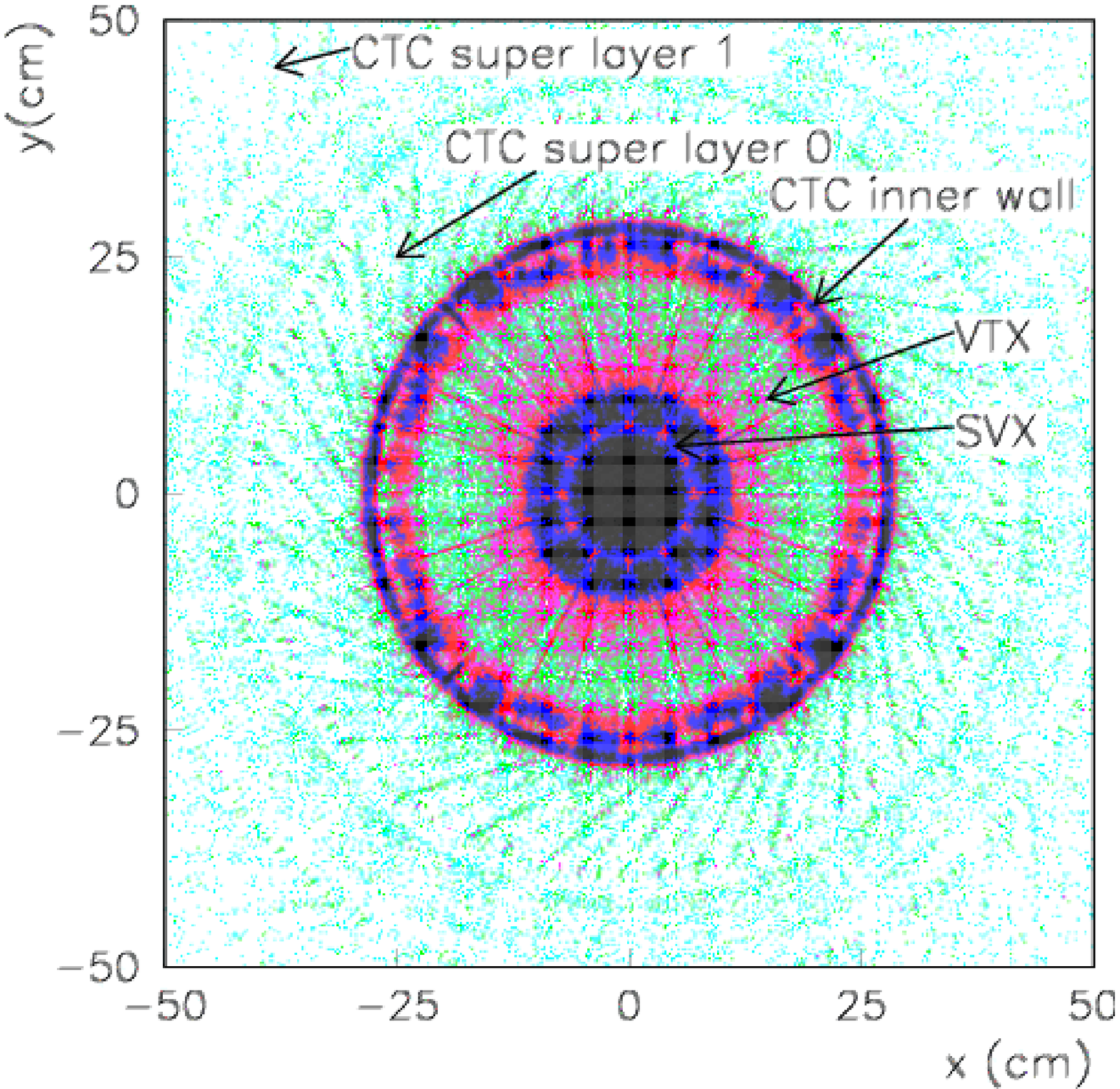}
\end{center}
\caption{Photon conversion vertex density  in the $x-y$ plane 
in the 74 pb$^{-1}$ of  CDF Run IB inclusive electron data.
The fine structure of the CDF tracking detectors
can be clearly resolved.}
\label{Fig:ConvCandxy}
\end{figure}

\clearpage

\begin{figure}
\begin{center}
\leavevmode
\epsfxsize=1.0\textwidth
\epsffile
{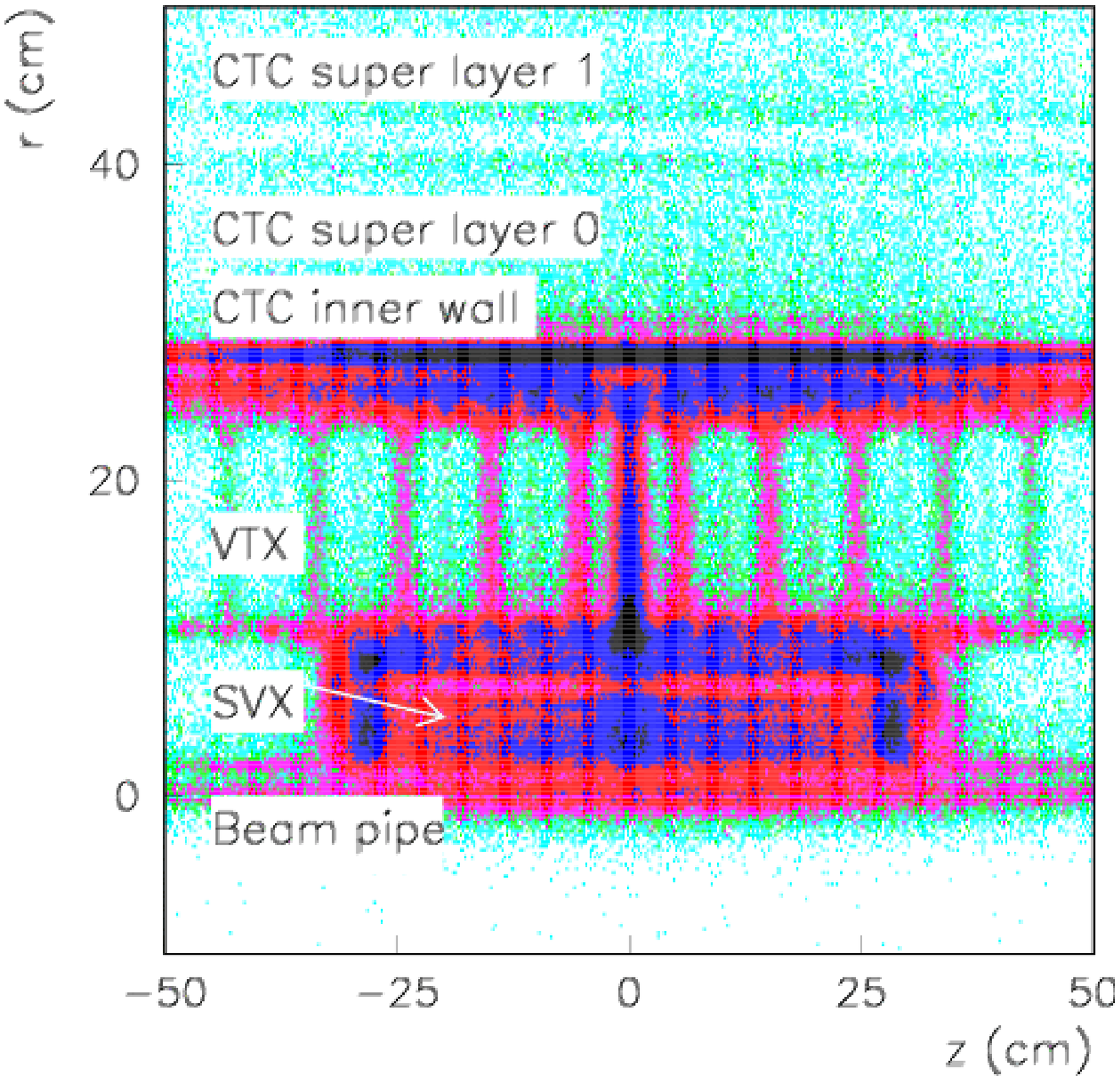}
\end{center}
\caption{Photon conversion vertex density  in the$r-z$ plane
 in the 74 pb$^{-1}$ of  CDF Run IB inclusive electron data.
The fine structure of the CDF tracking detectors
can be clearly resolved.}
\label{Fig:ConvCandrz}
\end{figure}

\clearpage

\begin{figure}
\begin{center}
\leavevmode
\epsfxsize=0.8\textwidth
\epsffile
{kstargamma_prd.epsi}
\end{center}
\caption{
Top:  $e^+e^-K^-\pi^+$ invariant mass distribution for
$\overline{B}_d^0 \rightarrow \overline{K}^{*0}(\rightarrow K^- \pi^+) \gamma(\rightarrow e^+ e^-)$
in the $74\;{\rm pb}^{-1}$ of CDF Run IB inclusive electron data.
Bottom:  corresponding $e^+e^-K^-$ invariant mass
distribution for the $B_u^-\rightarrow J/\psi(\rightarrow e^+ e^-) K^-$
reference decay.  There are $28.0\pm 5.8$ events after
background subtraction.
}
\label{Fig:KstarGammaConv}
\end{figure}

\clearpage

\begin{figure}
\begin{center}
\leavevmode
\epsfxsize=0.8\textwidth
\epsffile
{phigamma_prd.epsi}
\end{center}
\caption{
Top:  $e^+e^-K^+K^-$ invariant mass distribution for
 $\overline{B}_s^0 \rightarrow \phi(\rightarrow K^+ K^-) \gamma(\rightarrow e^+ e^-)$ in
                 the $74\;{\rm pb}^{-1}$ of CDF Run IB inclusive electron
                 data.
Bottom:  corresponding $e^+e^-K^-$ invariant mass
                 distribution for the 
                 $B_u^-\rightarrow J/\psi(\rightarrow e^+ e^-) K^-$
                 reference decay.  There are $35.0\pm 6.4$ events after
                 background subtraction.
}
\label{Fig:PhiGammaConv}
\end{figure}

\clearpage

\begin{figure}
\begin{center}
\leavevmode
\epsfxsize=0.8\textwidth
\epsffile
{lambdagamma_prd.epsi}
\end{center}
\caption{
Top:  $e^+e^-p\pi^-$ invariant mass distribution for
  $\Lambda_b^0 \rightarrow \Lambda(\rightarrow p \pi^-) \gamma(\rightarrow e^+ e^-)$ in
                 the $74\;{\rm pb}^{-1}$ of CDF Run IB inclusive electron
                 data.  
Bottom:  corresponding $e^+e^-K^-$ invariant mass
                 distribution for the 
                 $B_u^-\rightarrow J/\psi(\rightarrow e^+ e^-) K^-$
                 reference decay.  There are $24.0\pm 5.3$ events after
                 background subtraction.
}
\label{Fig:LambdaGammaConv}
\end{figure}

\clearpage

\begin{figure}
\begin{center}
\leavevmode
\epsfxsize=0.8\textwidth
\epsffile
{jpsiee_candidates_prd.epsi}
\end{center}
\caption{Dielectron invariant mass distribution of the $J/\psi \to e^+e^-$
candidates in the 74 pb$^{-1}$ of  CDF Run IB inclusive electron data. 
The number of the $J/\psi \to e^+e^-$ events obtained by fitting
the mass distribution to a function of 2 Gaussians and a polynomial
is $\sim$ 8000.}
\label{Fig:JpsieeCandidates}
\end{figure}
%
%


\end{document}